\documentclass[a4paper,12pt]{article} 

\usepackage{amsmath}
\usepackage{color}
\usepackage{amssymb}
\usepackage{bm}
\usepackage{graphicx}
\begin{document}

\begin{flushright}
{\bf OUJ-FTC-5}\\
{\bf OCHA-PP-360}\\
\end{flushright}

\vspace{15mm}
\begin{center}
{\Large\bf
A Formalism Useful to Study Beyond Squeezing in Non-linear Quantum Optics}
\vspace{7mm}

\baselineskip 18pt
{\bf The OUJ Tokyo Bunkyo Field Theory Collaboration}

\vspace{2mm}

Noriaki Aibara${}^{1}$, Naoaki Fujimoto${}^{2}$, So Katagiri${}^{3}$, Akio Sugamoto${}^{4, 5}$, Koichiro Yamaguchi${}^{1}$,
Tsukasa Yumibayashi${}^{6}$,

\vspace{2mm}

{\it ${}^{1}$Nature and Environment, Faculty of Liberal Arts, The Open University of Japan, Chiba 261-8586, Japan\\
${}^{2}$Department of Information Design, Faculty of Art and Design, Tama Art University, Hachioji, 192-0394 Japan \\
${}^{3}$Division of Arts and Sciences, The School of Graduate Studies,\\
The Open University of Japan, Chiba 261-8586, Japan\\
${}^{4}$Tokyo Bunkyo Study Center, The Open University of Japan (OUJ), \\
Tokyo 112-0012, Japan \\
${}^{5}$Ochanomizu University, 2-1-1 Ohtsuka, Bunkyo-ku, Tokyo 112-8610, Japan\\
${}^{6}$Department of Social Information Studies, Otsuma Women's University, 12 Sanban-cho, Chiyoda-ku, Tokyo 102-8357, Japan}

\end{center}

\vspace{5mm}
\begin{center}
\begin{minipage}{14cm}
\baselineskip 16pt
\noindent
\begin{abstract}
A general formalism is given in quantum optics within a ring cavity, in which a non-linear material is stored.  The method is Feynman graphical one, expressing the transition amplitude or S-matrix in terms of propagators and vertices.  The propagator includes the additional damping effect via the non-linear material as well as the reflection and penetration effects by mirrors.  Possible application of this formalism is discussed, in estimating the averaged number of produced photons, Husimi function, and the observables to examine beyond the squeezing mechanism of photons.
\end{abstract}

\end{minipage}
\end{center}

\newpage

\section{Introduction}

In optics, laser technology has developed extensively and is now able to elucidate the quantum behavior of photons.  That is, we can investigate quantum optics (QO) or quantum electrodynamics (QED) by using laser technologies.  See for example \cite{Grynberg}.  It is amazing that there exist transparent materials, having non-linear dielectric constants, such as Silica, BBO ($\beta$-Barium Borate), KTP (Potassium Titanyl Phosphate), LN (Lithium Niobate), KDP (Potassium Dideuterium Phosphate) {\it e.t.c.}.  For such non-linear materials the displacement vector (the electric flux density) $\bm{D}$ is not linearly proportional to the electric field $\bm{E}$, but we have 
 \begin{eqnarray}
&&\bm{D}=\varepsilon_0 (1+\chi^{(1)} ) \bm{E} + \bm{P}_{NL}, ~~\bm{H}=\bm{B}/\mu_0,  ~\mathrm{and} \\
&& (\bm{P}_{NL})_i=\varepsilon_0 \sum_{\mathrm{i, j, k, \cdots}} \left( \chi^{(2)}_{ijk} E_jE_k +  \chi^{(3)}_{ijkl} E_jE_kE_l +  \chi^{(4)}_{ijklm} E_jE_kE_lE_m +\cdots \right),
\end{eqnarray}
where $\bm{P}_{NL}$ is the non-linear polarization induced by the electric field, $\varepsilon_0$ and $\mu_0$ are the dielectric constant and the magnetic permeability of the vacuum, respectively, and $\chi^{(n)}$ $(n=1, 2, \cdots)$ are dielectric constants.

The tensor structure with indices $(i, j, \cdots)$ is important for an individual material, but here we ignore it for simplicity.  

Then, the action of QO or QED in the material reads
\begin{eqnarray}
&&S_{QO}=\frac{1}{2} \int dt~d^3x~  (\bm{E}\cdot \bm{D}-\bm{H} \cdot \bm{B}) \\
&&=\frac{1}{2} \int dt~d^3x~\left\{ \varepsilon_0 \left( E^2+ \chi^{(1)}E^2 + \chi^{(2)} E^3 + \chi^{(3)}E^4+  \chi^{(4)} E^5 +\cdots \right)-\bm{H} \cdot \bm{B}\right\} \label{action of QO}
\end{eqnarray}
which is non-linear in $E$.
In this way the non-linear dielectric constant $\chi^{(n)}$ introduces additional interactions to the QO (or QED) Lagrangian density, 
\begin{eqnarray}
 \mathcal{L}^{(n)}_{QO} = \frac{1}{2} \varepsilon_0 \chi^{(n)}E(t, \bm{x})^{n+1}. 
 \end{eqnarray}
The electric field is composed of the dominant classical part $E_{L}$, coming from the intense laser beam, and the quantum part $\hat{E}_{q}$ describing the creation and annihilation of photons:
\begin{eqnarray}
&&E(t, x) = E_{L}(t, \bm{x})+ \hat{E}_{q}(t, \bm{x}), ~~\mathrm{where}\\
&&E_L(t, \bm{x})=  E_L \; e^{-i(\omega_L t- \bm{k}_L \cdot \bm{x})} +  E_L^{\dagger}\; e^{i(\omega_L t- \bm{k}_L \cdot \bm{x})} , \\
&&\hat{E}_q(t, \bm{x})=-\partial_t \hat{A}_q(t, \bm{x})=\int d^3 k \sqrt{\frac{\hbar}{ (2\pi)^3 \varepsilon\; 2\omega}}(i \omega)\left(\hat{a}(\bm{k}) e^{-i(\omega t- \bm{k}\cdot \bm{x})} - \hat{a}(\bm{k})^{\dagger} e^{i(\omega t- \bm{k}\cdot \bm{x})} \right),~~~~~~\label{mode expansion of E}
\end{eqnarray}
where $\hat{\bm{A}}_q$ is the vector potential, $\hat{a}(\bm{k})$ and $\hat{a}(\bm{k})^{\dagger}$ are, respectively, annihilation and creation operators of photon with a wave vector $\bm{k}$, and $\varepsilon=\varepsilon_0(1+\chi^{(1)})$.\footnote{In reality, the vector potential, creation and annihilation operators have dependence on the direction of polarization vectors, but these tensor structure is also ignored here.} Here we assume a simple dispersion relation, $\omega=\omega(\bm{k})=c'|\bm{k}|$, with the light velocity $c'=1/\sqrt{\varepsilon \mu_0}$ in the non-linear material.

Now, we understand that the intense laser beam can create or annihilate photons by the interactions, $\mathcal{L}^{(n)}_{QO}~(n=2, 3, 4, \cdots)$, proportional to $\chi^{(n)}$, 
\begin{eqnarray}
\mathcal{L}^{(n)}_{QO}=\frac{1}{2}(n+1)\varepsilon_0 \chi^{(n)}E_L(t, \bm{x}) \hat{E}_q(t, \bm{x})^n.  \label{NL interaction in QO}
\end{eqnarray}
At each interaction point, both energy $(\hbar \omega)$ and momentum $(\hbar \bm{k})$ are compelled to conserve in the cavity. This is called the ``phase matching condition'' in QO, which implies that the interaction always satisfies the resonance condition in the cavity, and is amplified maximally.   As an example, if the laser beam with $(\omega_L, \bm{k}_L)$ create $n$ photons with  energy and momentum, $(\omega_1, \bm{k}_1), (\omega_1, \bm{k}_1), \cdots, (\omega_n, \bm{k}_n)$, then we have
\begin{eqnarray}
\omega_L=\omega_1+ \omega_2 + \cdots+ \omega_n, ~\mathrm{and}~~
\bm{k}_L=\bm{k}_1+ \bm{k}_2 + \cdots + \bm{k}_n.  \label{energy momentum conservation}
\end{eqnarray}
It is important to note that in the processes occurring in the ``ring cavity'', all the photons, including the laser photons, propagate in one-direction, and the motion in the opposite direction is prohibited.

The cavity called ``optical parametric oscillator (OPO)'' is an apparatus with one spacial dimension, such as a Fabry-P\'erot resonator and a ring resonator.  In Fabry-P\'erot resonator, there are both forward and backward motions by mirror reflection, while in the ring resonator, the motion is restricted to the one-way round trip.  Therefore, we will study this ring resonator in this paper, for simplicity.  In such one-dimensional one-way motion, the energy-momentum conservation in Eq.(\ref{energy momentum conservation}) is reduced solely to the energy conservation, 
\begin{eqnarray}
\omega_L=\omega_1+ \omega_2 + \cdots+ \omega_n,
\end{eqnarray}
where the momentum conservation is automatically guaranteed, since the magnitude of a momentum is proportional to its energy, $|\bm{k}|=\omega/c'$. 

The other possible processes are
\begin{eqnarray}
\omega_L+ (\omega_1+ \omega_2+ \cdots + \omega_m)=\omega_{m+1} +\cdots +\omega_{n},
\end{eqnarray}
which implies that the laser beam and the photons with energy $(\omega_1, \omega_2, \cdots, \omega_m)$ come into the cavity and annihilate, and the photons with $(\omega_{m+1}, \cdots, \omega_n)$ are created and go out from the cavity.

Therefore, QO in one-dimensional apparatus can be analyzed without referring to the momentum or the space coordinate, if the apparatus is uniform in sufficiently long length scales in between mirrors.

Now, it is reasonable to study QO in terms of the following non-linear quantum mechanics with time-dependent interactions,
\begin{eqnarray}
L=\frac{1}{2} \dot{x}^2 - \frac{1}{2} \omega_0^2 x^2- \frac{1}{2!} \lambda_2(t)x^2 - \frac{1}{4!} \lambda_4(t) x^4, \label{model Lagrangian}
\end{eqnarray}
We can include more higher-order terms such as $x^6, x^8, \cdots$.  Then, the quantum mechanics (QM) has non-linear interactions via the couplings $\lambda_n(t)$, given in the interaction picture as 
\begin{eqnarray}
 L^{(n)}=-\frac{1}{n!} \lambda_n(t) \hat{x}(t)^n, \label{NL interaction in QM}
\end{eqnarray}
where 
\begin{eqnarray}
\hat{x}(t)=\sqrt{\frac{\hbar}{2\omega_0}}\left(\hat{a} e^{-i \omega_0t}+\hat{a}^{\dagger} e^{i \omega_0t}\right)
\end{eqnarray}

In comparison of Eq.(\ref{NL interaction in QO}) in QO with Eq.(\ref{NL interaction in QM}) in QM, we understand immediately that in case of one spacial dimension, these equations become equivalent under the following correspondence:
\begin{eqnarray}
\lambda_n(t) \Leftrightarrow \varepsilon_0 \chi^{(n)} E_{L}(t), ~\mathrm{and}~~\hat{x}(t) \Leftrightarrow \hat{E}_q(t).
\end{eqnarray}  

The time dependent coupling $\lambda_2(t)$ in QM and the non-linear dielectric constant $\chi^{(2)}$ in QO equally generate the squeezing states, where the uncertainty for the ``coordinate'' is scaled up (down), while that of ``momentum'' is scaled down (up), keeping the product of them unchanged.  In QO, the vector potential $\bm{A}(t, \bm{x})$ and the electric field $\bm{E}(t, \bm{x})$ form a canonical conjugate pair, playing the role of  ``coordinate'' and ``momentum'', respectively.  

It is remarkable that the squeezing is observed in OPO, in which the Fabry-P\'erot resonator is implemented with a non-linear crystal of MgO:LiNb$\mathrm{O}_3$ stored inside the cavity \cite{Wu}.  The appearance of the squeezing states in the parametric down-conversion of laser photon $\gamma_L$ into two degenerate photons $\gamma_1$ and $\gamma_2$ ($\omega_L=\omega_1+\omega_2$ and $\omega_1=\omega_2$), is confirmed using the homodyne interferometer.

In this paper, we develop a simple formalism useful to study the phenomena, a more generally ``beyond the squeezing mechanism'', for example to see beyond the squeezing induced by the anharmonic coupling $\lambda_4(t)$ in QM or by the non-linear dielectric constant $\chi^{(4)}$ in QO.  We consider a one-dimensional ring cavity, having a non-linear optical material and mirrors.  See (Figure\ref{apparatus}). 
For this purpose, we develop a perturbation theory in which the Hamiltonian is separated into the quasi-free Hamiltonian and the real interaction Hamiltonian.  The former includes the effects of dispersion and absorption by the non-linear optical material and the mirrors.  The intense laser light is prepared, which forms a classical light far above the threshold of laser oscillation.  To formulate the perturbation theory consistently, the Gaussian averaging $e^{-(|\omega_k^L|-\omega_R)^2/2\delta^2}$ is required over the laser energy $\omega_k^L$ around the resonance energy $\omega_R$ at $k$-th interaction point.  This averaging tames the singularities (mass singularities), which are inevitable when both the conservations of energy and momentum are compelled to hold so that the phase matching conditions hold. 

A lot of works have been done so far on the non-linear quantum optics, based on the input-putout theory, the Fokker-Planck equation, and the QED in non-uniform dielectric matter \cite{key-49, key-4}, \cite{Fokker-Planck}, \cite{QED}.  We prepare Appendices, A, B, and C corresponding to the three issues, in which we give comments on the relation of this paper to the issues. 

The more general case of beyond the squeezing, having the higher-order interactions, was studied separately in \cite{Katagiri}, by referring to the Virasoro algebra, in which a general formula is derived, which gives the number of particles produced, under a condition of the number of particles produced is large.  

Beyond the squeezing is also called ``generalized squeezing'' \cite{generalized squeezing}. These generalized squeezed states are given by applying the following operator to the vacuum $(k \ge 3)$,
\begin{equation}
\hat{U}_{k}=\exp\left(z\hat{a}^{k}+z^{*}\hat{a}^{\dagger k}\right)
\end{equation}
which are naturally conceivable from the analogy of coherent states $(k=1)$,
\begin{equation}
\hat{U}_{1}=\exp\left(z\hat{a}+z^{*}\hat{a}^{\dagger}\right),
\end{equation}
and the squeezed states $(k=2)$,
\begin{equation}
\hat{U}_{2}=\exp\left(z\hat{a}^{2}+z^{*}\hat{a}^{\dagger2}\right).
\end{equation}
It is, however, controversial, whether the vacuum expectation value of the generalized squeezed state $\langle 0 |\hat{U}_k^{\dagger} \hat{U}_k|0\rangle$ $(k \ge 3)$ is divergent or convergent.  See Appendix D for the detail. 

In the next section, a ring resonator and its resonance condition are discussed.  Transition amplitude for ring resonator is formulated in Section 3.  In Section 4, the propagator and the vertex are explicitly given for QO inside ring resonator in the presence of NL-material, which are necessary items to estimate the transition amplitude. Transition amplitude is written using the propagators and the vertex functions in Section 5, so that our formalism can be applicable to higher-order calculations in QO in the cavity.  
The final section is devoted to conclusion and discussion; where possible application of the formalism developed in this paper is discussed, such as in estimating averaged number of produced photons, Husimi function and the observables to examine beyond the squeezing of photons.  

Appendices A, B, C, and D survey , respectively, the input-output theory and observation of squeezed state in optical cavity, the Fokker-Planck equation (including the condition of laser light be classical), QED in non-uniform dielectric matter, and a controversial issue on generalized squeezing.

\section{Ring Resonator for Optical Parametric Oscillation}

We adopt a ``ring resonator" as the apparatus to observe the beyond squeezing (generalized squeezing) by Optical Parametric Oscillation (OPO), which is depicted in (Figure 1). 

{\begin{figure}[h]
\centering
\includegraphics[width=140mm,clip]{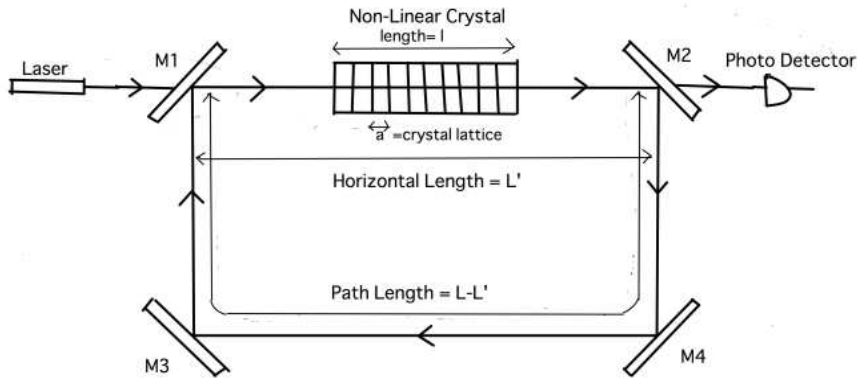}
\caption{A ring resonator as an apparatus to observe beyond the squeezing by Optical Parametric Oscillation (OPO).}
\label{apparatus}
\end{figure}}

The ring resonator, confined in a cavity, consists of flour mirrors, $M_1-M_4$.  The $M_1$ is the half-mirror, through which the input laser beam comes in from the left to the cavity, and $M_2$ is the output half-mirror, through which the photons go out and are detected by the photo-diode (PD) located outside the cavity.  The remaining $M_3$ and $M_4$ are ordinal mirrors.  The non-linear (NL) crystal of length $\ell$ is placed in between $M_1$ and $M_2$.  The path length of light from $M_1$ to $M_2$ is $L'$ and $L$ is the length of a round trip in the resonator.  The ring resonator extends in one spacial direction, along which the coordinate $z$ is taken.  Choosing $z=0$ in the center of the crystal, any path length is measured by the coordinate $z$ from the origin, clock-wisely along the direction of propagation of light.  

The crystal with non-linear (NL) dielectric constants is assumed to have crystal planes perpendicular to $z$.  The planes are separated regularly by a separation of a lattice constant of the crystal, $a_{crys}$.  

We consider the laser beam to be an intense ``classical wave'' in this paper.  This assumption is not bad, since above the threshold of laser oscillation (the pumping rate dominates over the loss in the cavity), laser photons form a coherent state, so that the laser light can be considered as a classical electric field $E_L(t, z)$ with a well-defined phase without fluctuations \cite{Grynberg}. The validity of treating laser beam as classical one at well above the threshold is discussed in Appendix B, using Fokker-Planck equation.  The oscillation of this electric field of the laser in the perpendicular direction to $z$, induces the oscillation of  electrons on the crystal planes in the same perpendicular direction, from which photons are generated via the non-linear interaction in Eq.(\ref{NL interaction in QO}).

First, we will estimate the resonance condition of the cavity.  We denote $r_i$ the reflection coefficient of the i-th mirror, for the amplitude.  Then, the corresponding transmission coefficient is $\sqrt{1-r_{i}^2}$.  

The electric field of light with angular frequency $\omega$ can be written as
\begin{eqnarray}
E(t, z)=E\;e^{-i\omega \left(t- \frac{z}{c(z)} \right)- \gamma(z) z},
\end{eqnarray}
where the light velocity $c(z)$ can depend on the position $z$ to pass through; we choose $c$ in the vacuum and $c'=c/n_{\mathrm{crys}}$ in the crystal with the refraction coefficient $n_{\mathrm{crys}}$, while $\gamma(z)$ is the decay rate per unit length at $z$; we choose $\gamma(z)=0$ in the vacuum and $\gamma(z)=\gamma'$ in the crystal.

Consider the light wave starting from $z$ at $t$.  It undergoes round trips inside the cavity arbitrary times, being reflected by the mirrors, so that its electric field obtains an extra factor $K(\omega L/c)$, which includes both the phase change and the attenuation:
\begin{eqnarray}
K(\omega L/c)= \frac{1}{1- (r_1r_2r_3r_4) e^{i \phi(\omega L/c)- \gamma' \ell}},
\end{eqnarray} \label{K}
where 
\begin{eqnarray}
\phi(\omega L/c)=\frac{\omega}{c}(L-\ell) + \frac{\omega}{c'} \ell. \label{phi}
\end{eqnarray}

So, the strength of the electric field is proportional to $|K|^2$, which gives a resonance behavior:
\begin{eqnarray}
|K(\omega L/c) |^2= \frac{1}{\left\{1- r e^{- \gamma' \ell}\right\}^2 + 4  r e^{- \gamma'  \ell} \sin^2 \left\{\frac{1}{2}\phi(\omega L/c)\right\}},
\end{eqnarray}
where $r=r_1r_2r_3r_4$. 

If the resonance point at which $|K|^2$ takes the maximum value, is denoted by $\omega_R$ or its frequency $\nu_R$, we have the resonance condition as follows: 
\begin{eqnarray}
\frac{\phi(\omega_R L/c)}{2\pi}=\left(\frac{\omega_R}{\omega_{FSR}}\right) \left(1+ \frac{\ell}{L}(n_{\mathrm{crys}}-1)\right) = \mathrm{integer}, \label{resonance condition}
\end{eqnarray}
where $\omega_{FSR}=2\pi c/L$ is the angular frequency of a fictitious wave with wave length $L$, which is called the ``free streaming range", since adjacent resonances are separated by $\omega_{FSR}$.  The full width at half maximum $(\omega_R/\omega_{FSR})_{FWHM}$ of the resonance is roughly
\begin{eqnarray}
\left(\frac{\omega_R}{\omega_{FSR}}\right)_{FWHM} \approx \frac{|1-re^{-\gamma' \ell} |}{\sqrt{r e^{-\gamma' \ell}}} \ll 1, \label{FWHM}
\end{eqnarray}
since $r \approx 1$.  

Therefore, we can tune $L$ very precisely so that the waves resonate and are enhanced in the cavity.  The resonance is very sharp, and $\omega_R$ can be precisely selected.  It is important to note here that if $\omega_1$ satisfies the minimum resonance condition:
\begin{eqnarray}
\frac{\phi(\omega_1 L/c)}{2\pi}=\left(\frac{\omega_1}{\omega_{FSR}}\right) \left(1+ \frac{\ell}{L}(n_{\mathrm{crys}}-1)\right) = 1, 
\end{eqnarray}
then, $2\omega_1, 3\omega_1, 4\omega_1, \cdots$ also satisfy the resonance condition Eq.(\ref{resonance condition}) by choosing the integer as $2, 3, 4, \cdots$, if the refraction index $n_{crys}$, (generally depends on the frequency $\omega$), is constant for these waves.  The FWHM in Eq.(\ref{FWHM}) depends on the waves, since it depends on the $r$ and $\gamma'$ whose dependency on $\omega$ may not be ignored.   

We consider a crystal, having NL dielectric constants $\chi^{(2)}$ and $\chi^{(4)}$.  Generalization to more general cases with $\chi^{(i)}~(i=3,  5, 6, \cdots)$ is also possible.\footnote{The ring resonator which we are considering is especially suited for observing the parity violation effects from $\chi^{(i)}, i=$odd integer. See for example \cite{Fan}, in which the ring resonator is examined to see the parity violation effect of neutrinos as dark matter, without QED background.  On the other hand, the forward and backward motions in Fabry-P\'erot cavity wash out the parity violation effects.}

Now, we choose the laser frequency to be $\omega_L=\omega_0$, then the down-converted photons with frequencies, $\omega_L/4=\omega_1$ and $\omega_L/2=\omega_2=2\omega_1$, can also resonate and survive in the cavity.\footnote{To input the laser beam smoothly into the ring cavity and to extract the output photons effectively from it, some devise may be necessary.  Referring to the gravitational wave detector \cite{LIGO}, a way is to prepare two more cavities, the laser input cavity (LIC) and the signal extraction cavity (SEC). The former cavity (LIC) consists of the mirror $M_1$ and mirrors in front and behind of $M_1$ along the direction of laser beam, while the latter cavity (SEC) consists of the mirror $M_2$ and mirrors in front and behind of $M_2$ along the path to the photo-detector.  The LIC can be designed almost transparent for $\omega_L$ but very reflective for $\omega_1$ and $\omega_2$, while the SEC can be designed very reflective for $\omega_L$ but almost transparent for $\omega_1$ and/or $\omega_2$.  If necessary, the advanced technique such as modulation and demodulation of amplitude (AM) and of phase (PM) can be used. }

In the cavity, the momenta of photons should always be parallel to the longitudinal direction.  That is, if the photon in the cavity is scattered by the crystal and loses energy or alters the direction of momentum, then this photon can not exist in the cavity and disappears; the damping rate $\gamma'$ comes from these scatterings.

\section{Transition amplitude for Ring Resonator} 
In this section, we will study the transition amplitude in QO, which describes the creation and annihilation of photons from a non-linear crystal, induced by the laser.  In this QO inside the resonator, the Hamiltonian can be divided into two parts; the first part is denoted as $\hat{H}_0$, which is time dependent and not free but is ``quasi-free'', including the effect of damping in the cavity and the reflection and penetration by the mirrors. The second part $\hat{H}_1$ gives the real interaction (creation and annihilation) of photons by the non-linear crystal.  Thus, the Hamiltonian which we are going to study is time dependent, even in the ``Shr\"odingier picture'':
\begin{eqnarray}
\hat{H}(t)=\hat{H}_0(t)+\hat{H}_1(t).
\end{eqnarray}

We will call $\hat{H}_0$ the ``quasi-free Hamiltonian'', and $\hat{H}_1$ the ``real interaction Hamiltonian''.

We will mimic the process of picture changing from the ``Schr\"odinger picture'' to the ``interaction picture'', by treating $\hat{H}_0$ as if it was the free Hamiltonian.  

If we introduce $\hat{U}_0(t)$, $U$-matrix or the time evolution operator for the quasi-free Hamiltonian $\hat{H}_0(t)$, it satisfies
\begin{eqnarray}
i\hbar \; \frac{\partial}{\partial t}\hat{U}_0(t) = \hat{H}_0(t) \hat{U}_0(t).
\end{eqnarray}
Explicitly, we have 
\begin{eqnarray}
\hat{U}_0(t)=T e^{\int_0^t dt'\hat{H}_0 (t')/i\hbar}.
\end{eqnarray}

In this ``Schr\"odingier picture", the equations of motion of wave function $\psi(t)_S$ and operator $\hat{O}_S$ are given by
\begin{eqnarray}
 i \hbar \; \frac{\partial}{\partial t}\psi(t)_S = \hat{H}(t) \psi(t)_S, ~\mathrm{and}~~ i\hbar \; \frac{\partial}{\partial t}\hat{O}_S=0,
 \end{eqnarray}

After moving to the ``interaction picture" we have 
\begin{eqnarray}
i \hbar \; \frac{\partial}{\partial t} \hat{O}_I =  [ \hat{O}_I, \hat{H}_0(t)_S], ~~\mathrm{and}~~i \hbar \; \frac{\partial}{\partial t} \psi(t)_I = \hat{H}_1(t)_I \psi(t)_I, \label{wave equation in I picture} 
\end{eqnarray}
where
\begin{eqnarray}
\psi(t)_I= \hat{U}^{-1}_0(t) \psi(t)_S, ~\mathrm{and}~\hat{H}_1(t)_I=\hat{U}_0(t)^{-1} \hat{H}_1(t)\hat{U}_0(t).
\end{eqnarray}
Then, in the time evolution of the ``interaction picture", the quasi-free motion is subtracted from that of the ``Schr\"odingier picture''. The motion of laser light and photons, before and after the real interaction in the cavity occurs, is quasi-free, so that in these periods the wave function in the ``interaction picture" does not evolve.  

 The transition amplitude of QO in the ring resonator can be expressed iteratively as usual, by solving the time evolution of wave function in the ``interaction picture".  We will use the number representation, in which the sate is expanded in $|n\rangle$ having $n$ photons.  Then, the Schr\"odingier equation in the interaction picture, $i\hbar \frac{\partial}{\partial t} |n, t\rangle = \hat{H}_1(t)_I |n, t\rangle$, can be solved, if we choose the initial condition $| n, -\infty \rangle=|m\rangle$, as follows 
\begin{eqnarray}
&&|n, t \rangle= |m \rangle + \frac{1}{i \hbar}\int_{-\infty}^{t} dt_1 \; |n \rangle \langle n| \hat{H}_1(t_1)_I |m \rangle \nonumber \\
&&~~+ \left(\frac{1}{i \hbar}\right)^2 \int_{-\infty}^{t} dt_1 \int_{^\infty}^{t_1} dt_2 \; |n \rangle \langle n| \hat{H}_1(t_1)_I |m_1 \rangle \langle m_1| \hat{H}_1(t_2)_I |m \rangle + \cdots~~~ \\
&&=|m \rangle + \frac{1}{i \hbar}\int_{-\infty}^{t} dt_1 \; |n \rangle \langle n |\hat{U}_0(t)^{-1}|n' \rangle \nonumber \\
&&~~~~~~~~~~~~~\times \langle n' | \hat{U}_0(t, t_1) |m_1 \rangle \langle m_1| \hat{H}_1(t_1) |m_2 \rangle \langle m_2 |\hat{U}_0(t_1, -\infty) | m' \rangle \times \langle m' | \hat{U}_0(-\infty) | m \rangle \nonumber \\
&& ~~+ \left(\frac{1}{i \hbar}\right)^2 \int_{-\infty}^{t} dt_1 \int_{^\infty}^{t_1} dt_2 \; |n \rangle \langle n |\hat{U}_0(t)^{-1}|n' \rangle \nonumber \\
&&~~\times \langle n'| \hat{U}_0(t, t_1) |m_1 \rangle \langle m_1| \hat{H}_1(t_1) |m_2 \rangle \langle m_2| \hat{U}_0(t_1, t_2) |m_3 \rangle\langle m_3| \hat{H}_1(t_2) |m_4 \rangle \langle m_4 |\hat{U}_0(t_2, -\infty) | m' \rangle \nonumber \\
&&~~\times \langle m' | \hat{U}_0(-\infty) | m \rangle + \cdots ,
\end{eqnarray}
where $\hat{U}_0(t_1, t_2) \equiv \hat{U}_0(t_1)\hat{U}_0(t_2)^{-1}$ is a transition amplitude from $t_2$ to $t_1$ by the ``quasi-free Hamiltonian $\hat{H}_0$".  When $t$ is finite, then the factor $\langle n |\hat{U}_0(t)^{-1}|n' \rangle$ should be carefully estimated, but in our case, we will take $t=-\infty$, and consider the transition amplitude from $t=-\infty$ to $t=\infty$, that is the S-matrix, we may choose
\begin{eqnarray}
\langle n |\hat{U}_0(\infty)^{-1}|n' \rangle=\delta_{nn'}, ~\langle m' | \hat{U}_0(-\infty) | m \rangle=\delta_{mm'}.
\end{eqnarray}

Next, we have to discuss an important point that QO is essentially a field theory, having spacial degrees of freedom, $\bm{x}$. Then, the Hamiltonian is given as the integral over $\bm{x}$ of the ``Hamiltonian density $\mathcal{H}(t, \bm{x})$'', that is
\begin{eqnarray}
\hat{H}_0(t) = \int d\bm{x} \; \mathcal{H}_0(t, \bm{x}), ~~\mathrm{and}~~\hat{H}_1(t) = \int d\bm{x} \; \mathcal{H}_1(t, \bm{x}).
\end{eqnarray}
Thus, the transition amplitude in QO can be written as follows:
\begin{eqnarray}
&&|n, t \rangle=|m \rangle + \frac{1}{i \hbar}\int_{-\infty}^{t} dt_1\int d \bm{x}_1 \; |n \rangle \langle n |\hat{U}_0(t)^{-1}|n' \rangle \nonumber \\
&&\times \langle n' | \hat{U}_0(t, t_1) |m_1 \rangle \langle m_1 | \mathcal{H}_1(t_1, \bm{x}_1) |m_2 \rangle \langle m_2 | \hat{U}_0(t_1, -\infty) |m' \rangle \nonumber \\
&&\times \langle m' | \hat{U}_0(-\infty) |m \rangle  \nonumber \\
&& + \left(\frac{1}{i \hbar}\right)^2 \int_{-\infty}^{t} dt_1 \int d\bm{x}_1 \int_{^\infty}^{t_1} dt_2 \int d\bm{x}_2 \; |n \rangle \langle n |\hat{U}_0(t)^{-1}|n'  \rangle  \nonumber \\
&& \times \langle n' | \hat{U}_0(t, t_1) |m_1 \rangle \langle m_1 | \mathcal{H}_1(t_1, \bm{x}_1) |m_2 \rangle \langle m_2 |\hat{U}_0(t_1, t_2) |m_3 \rangle \nonumber \\
&&\times \langle m_3 | H_1(t_2, \bm{x}_2) |m_4 \rangle \langle m_4 |\hat{U}_0(t_2, -\infty) |m' \rangle \times  \langle m' |\hat{U}_0(-\infty) |m \rangle \nonumber \\
&&+ \cdots.
\end{eqnarray}
 
\section{Propagator and Vertex inside Ring Resonator containing Non-linear Material}

We first examine a simple case of quantum optics (QO) inside the uniformly distributed non-linear material, which was discussed a bit in Introduction.  The Lagrangian density in this case is Eq.(\ref{action of QO}),
 \begin{eqnarray}
&&\mathcal{L}_{QO}=\frac{1}{2}  (\bm{E}\cdot \bm{D}-\bm{H} \cdot \bm{B})=\mathcal{L}_0+\mathcal{L}_1,  
\end{eqnarray}
where 
\begin{eqnarray}
\mathcal{L}_0=\frac{1}{2} \varepsilon \left(\bm{E}^2 - \frac{1}{\varepsilon\mu_0} \bm{B}^2 \right),  ~~\mathrm{and}~~\mathcal{L}_1= \frac{1}{2} \varepsilon_0 \sum_{n \ge 2} \chi^{(n)} (\bm{E})^{n+1}.
\end{eqnarray}
The quadratic terms are combined into a quasi-free Lagrangian with a new parameter $\varepsilon=\varepsilon_0(1+\chi^{(1)})$, the dielectric constant of the material.

In terms of vector potential $\bm{A}(t, \bm{x})$ and electric field $\bm{E}=-\dot{\bm{A}}$, the conjugate momentum of the former is given by $\bm{\Pi}_A=-\varepsilon \bm{E}=\varepsilon \dot{\bm{A}}$, and $\bm{B}=\bm{\nabla} \times \bm{A}$, so that the Hamiltonian density $\mathcal{H}$ reads 
\begin{eqnarray}
\mathcal{H}=\mathcal{H}_0+ \mathcal{H}_1,
\end{eqnarray}
where
\begin{eqnarray}
\mathcal{H}_0=\frac{1}{2} \varepsilon \left(\bm{E}^2 + \frac{1}{\varepsilon\mu_0} \bm{B}^2 \right),  ~~\mathrm{and}~~\mathcal{H}_1= \frac{1}{2} \varepsilon_0 \sum_{n \ge 2} n \chi^{(n)} (\bm{E})^{n+1}.
\end{eqnarray}

The quasi-free Hamiltonian, in this case, is a free Hamiltonian in the medium, where $c'=c/n=c/\sqrt{\varepsilon\mu_0}$ ($n$: the refraction constant) is the light velocity in the medium.

Now, we are ready to study QO inside the ring resonator.

\subsection{Mode Expansion}
In QED, any mode having an arbitrary wave vector $\bm{k}$ have to be considered, but in QO, only a few modes, having specific values of wave vectors, have to be considered due to the resonance condition of the cavity.  In this situation, it is better to formulate QO in terms of discretized coordinates and momenta.  Let us denote discretized coordinates with equal spacing $a$ by $\bm{x}_n$, and the discretized (angular frequency $\omega$, wave vector $\bm{k}$) with equal spacing by $(\omega_m, \bm{k}_m)$.  The total number of them are equally $N (\gg 1)$, that is, $(n, m)=1, 2, \cdots, N$, and $Na=L$ is the total length of the system.  Then, the spacing of the wave vector is $2\pi/L$ as usual.  Corresponding to the mode expansion of $\bm{E}$ in Eq.(\ref{mode expansion of E}), the mode expansion of the vector potential $\bm{A}$ and its canonical conjugate $\bm{\Pi}_A$, are as follows\footnote{The reason why the dielectric constant $\varepsilon$ appears in the mode expansion is that the conjugate momentum defined by $\hat{\Pi}_A=\varepsilon \partial_t \hat{A}$ in the medium.}:
\begin{eqnarray}
&&\hat{A}_q(t, \bm{x})=\int d^3 k \sqrt{\frac{\hbar}{ (2\pi)^3 \varepsilon\; 2\omega}}\left(\hat{a}(\bm{k}) e^{-i(\omega t- \bm{k}\cdot \bm{x})} + \hat{a}(\bm{k})^{\dagger} e^{i(\omega t- \bm{k}\cdot \bm{x})} \right), \\
&&\hat{\Pi}_{A, q}(t, \bm{x})=\varepsilon \int d^3 k \sqrt{\frac{\hbar}{ (2\pi)^3 \varepsilon\; 2\omega}}(-i \omega) \left(\hat{a}(\bm{k}) e^{-i(\omega t- \bm{k}\cdot \bm{x})} - \hat{a}(\bm{k})^{\dagger} e^{i(\omega t- \bm{k}\cdot \bm{x})} \right),
\end{eqnarray}
giving the non-vanishing commutation relations,
\begin{eqnarray}
[\hat{A}_q(t, \bm{x}), \hat{\Pi}_{A, q}(t, \bm{x}')]=i \hbar \; \delta^{(3)} (\bm{x}-\bm{x}'), ~\mathrm{or~equivalently},\;  [\hat{a}(\bm{k}), \hat{a}(\bm{k}')^{\dagger}]=\delta^{(3)} (\bm{k}-\bm{k}').
\end{eqnarray} 

If we restrict the problem to QO in one-dimensional cavity, the discretized version of operators can be introduced by\footnote{In case of three spacial dimensions, the square roots should be replaced by $a^{\frac{3}{2}}$ or $\left(\frac{2\pi}{L}\right)^{\frac{3}{2}}$.} 
\begin{eqnarray}
&&\sqrt{a}\;  \hat{A}_q(t, \bm{x})= \hat{A}(t, \bm{x}_n), ~\sqrt{a}\;  \hat{\Pi}_{A, q}(t, \bm{x})= \hat{\Pi}_{A, q}(t, \bm{x}_n), \label{discrete coordinate}\\
&&\sqrt{\frac{2\pi}{L}}\; \hat{a}(\bm{k})=\hat{a}(\bm{k}_m), ~\sqrt{\frac{2\pi}{L}}\;  \hat{a}(\bm{k})^{\dagger}= \hat{a}(\bm{k}_m)^{\dagger}. \label{discrete wave vector}
\end{eqnarray} 

Then, the commutation relations are expressed with Kroneker's delta as follows:
\begin{eqnarray}
[\hat{A}_q(t, \bm{x}_n), \hat{\Pi}_{A, q}(t, \bm{x}_{n'})]=i \hbar \; \delta_{nn'}, ~\mathrm{or~equivalently},\;  [\hat{a}(\bm{k}_m), \hat{a}(\bm{k}_{m'})^{\dagger}]=\delta_{mm'}.
\end{eqnarray}

Thus, the discretized version of the vector potential and the electric field are given by
\begin{eqnarray}
&&\hat{A}_q(t, z_n)=\sum_{m=1}^N \sqrt{\frac{\hbar}{ N \varepsilon\; 2\omega_m}}\left(\hat{a}_m e^{-i(\omega_m t- k_mz_n)} + \hat{a}^{\dagger}_m e^{i(\omega_m t- k_mz_n)}  \right), \\
&&\hat{E}_q(t, z_n)=\sum_{m=1}^N \sqrt{\frac{\hbar}{ N \varepsilon\; 2\omega_m}}(i \omega_m) \left(\hat{a}_m e^{-i(\omega_m t- k_mz_n)} - \hat{a}^{\dagger}_m e^{i(\omega_m t- k_mz_n)}  \right).
\end{eqnarray}
These equations are valid for any spacial dimension, if we consider $N$ to be the total number of discretized points in coordinate space and wave vector space.
Similarly, the classical part of laser in Eq.(\ref{mode expansion of E}), is renormalized, when using the discrete coordinates, 
\begin{eqnarray}
\sqrt{a} E_L(t, z)= E_L(t, z_n)~\mathrm{with}~\begin{cases}
E_L(t, z)=  E_L \; e^{-i\omega_L (t-z/c(z))} +  E_L^{\dagger}\; e^{i\omega_L (t-z/c(z))}, \\
E_L(t, z_n)=  E_L \; e^{-i\omega_L (t-z_n/c(z))} +  E_L^{\dagger}\; e^{i\omega_L (t-z_n/c(z))}.
\end{cases}
\end{eqnarray}

In the above, the discrete sum over coordinates and that over wave vectors appear.  These two discretizations are correlated, but it is convenient to convert, while keeping the discrete sum over modes, the discrete sum to the coordinate integration, by using $\sum_n= \int dz/a$. This is valid since $a$ is small compared to the wave length.  The discretization of modes appears not from the correspondence between $a$ and $L$, but from the resonance condition.  In the following, we adopt this hybrid expression, that is, coordinate is continuous, while the wave vector is discrete.  

In our ring cavity, only two specific modes of photons, having angular frequencies, $\omega_1=\omega_L/4$ and $\omega_2=\omega_L/2$ are relevant.  

Therefore, in the hybrid expression, the classical part of electric field from laser is the same as Eq.(\ref{mode expansion of E}), but the quantum part of the electric field is chosen as follows, namely 
\begin{eqnarray}
&&E_L(t, z)=  E_L \; e^{-i\omega_L (t-z/c(z))} +  E_L^{\dagger}\; e^{i\omega_L (t-z/c(z))}, ~\mathrm{but} \\
&&\hat{E}_q(t, z)=i\sum_{i=1,\; 2} \sqrt{\frac{\hbar \omega_i}{ 2L\varepsilon (z)}}\left(\hat{a}_i e^{-i(\omega_i(t- z/c(z))} - \hat{a}^{\dagger}_i e^{i\omega_i (t- z/c(z))}  \right),
\end{eqnarray} 
where the system size $L$ remains, and $c(z), \; \varepsilon(z)$ are the light velocity and the dielectric constant at a position $z$, respectively.  It is natural to choose $L$ be a total path length of the ring resonator.  

\subsection{Propagator}
The ``propagator" in the number representation is given as
\begin{eqnarray}
\langle n |\hat{U}_0(t, t') |n' \rangle=\langle n |\hat{U}_0(t)\hat{U}_0(t')^{-1} |n' \rangle.
\end{eqnarray}
Here, $\hat{U}(t, t')$ satisfies Eq.(\ref{wave equation in I picture}), so that it is true to give the phase change of wave in time, $e^{-\frac{i}{\hbar} \hat{H}_0 (t-t')}=(e^{-i \omega (t-t')})^n$ for the $n$ photon state, but that is not all.  The propagation of laser beam and of produced photons move clock-wisely inside the cavity, inside which non-linear material and the mirrors exist, so that the wave changes position dependently.  In our case with one spacial dimension, the position $z$ is determined by the time $t$.  Therefore, we will write the propagator as a time-ordered product, being defined piece-wisely.  As an example, let us consider a photon with the frequency $\omega$ propagates first from $(t_0, z_0)$ to $(t_1, z_1)$ inside the medium with velocity $c'$, next propagates in the vacuum till $(t_2, z_2)$, and it is reflected afterwards by a mirror with a reflection constant $r$.  After that it is assumed to propagates in the vacuum, and finally arrives at $(t_3, z_3)$.  The corresponding propagator can be  
\begin{eqnarray}
U(t_3, t_0)= U(t_3, t_2, \mathrm{vac}) \times r \times U(t_2, t_1; \mathrm{vac}) \times U(t_1, t_0; \mathrm{medium}),
\end{eqnarray}
where $U(t_1, t_0; \mathrm{vac})=e^{-i\omega\{(t_1-t_0)-(z_1-z_0)/c\}}$, but in the medium, damping occurs with rate $\gamma'$ per length, so that $U(t_1, t_0; \mathrm{medium})=e^{-i\omega\{(t_1-t_0)-(z_1-z_0)/c'-\gamma'(z_1-z_0)}$. The derivation of the propagator is the same both in the classical wave and the quantum wave.\footnote{If we wish to derive more rigorously the propagators, we can do it, by considering the Hilbert space of photons as a product of that inside the cavity (or that of the system) $\mathcal{H}_s$ and that outside the cavity (or that of the heat bath) $\mathcal{H}_b$, namely, $\mathcal{H}=\mathcal{H}_s \otimes \mathcal{H}_b$.  Since the attenuation caused by $\gamma'$ and $r$ is the effect of escape of photons from the system to the heat bath.  If we trace out the heat bath and concentrate to the system, we have the effect of attenuation, caused by the crystal, and the reflection and penetration via mirrors. The rigorous result does not, however, differ from that given in the above.  (See Appendix A and B for the details.)}

Now, we can understand that the propagator of $n$ photons is
\begin{eqnarray}
\langle n |\hat{U}_0(t, t') |n' \rangle \equiv \delta_{nn'} D_n(t, z; t', z'), ~\mathrm{and}~ D_n(t, z; t', z')= \langle 1 |\hat{U}_0(t, t') |1 \rangle^n \equiv D(t, z; t', z')^n,
\end{eqnarray}
where $1/n!$ appeared from the normalization of the state is cancelled by the degeneracy $n!$ which comes from the ways of connecting $n$ (initial and final) photons by propagators.

The propagator of a single photon, $D(t, z; t, z')$, describes the transition amplitude of state by the quasi-free Hamiltonian.  Its derivation is the same as that of the coefficient $K(\omega L/c))$ in Eq.(\ref{K}), including the round trips.  Difference between $K$ and the propagator $D$ is that the starting point $(t', z')$ and the arriving point $(t', z')$ are equal in the former, but different in the latter.  

There are a number of propagators:

1) ``propagator" from a crystal plane $C'$ to another crystal plane $C$:

\begin{eqnarray}
D^{\omega}_{C; C'}(t, z; t', z') =\frac{e^{-i \omega \{(t-t') -(z-z')/c'\}-\gamma'|z-z'|}}{1- \left(r_{\omega} e^{i \phi (\omega L/c)-\gamma'\ell}\right) } \equiv e^{-i \omega \{(t-t') -(z-z')/c'\}} K^{\omega}_{C;C'}(z, z'). \label{K1}
\end{eqnarray}
Here, $\phi(\omega L/c)$ is given in Eq.(\ref{phi}) in the previous section.
This is applicable to $\omega=\omega_1$, or $\omega_2(=2\omega_1)$.  If the $n$ photons propagate, then the factor becomes $(D^{\omega}_{C; C'}(t, z; t', z'))^n$.  

2) ``propagator" for the laser light:

The $\hat{U}_0(t, t')$ is purely quantum object, giving the transformation of operators from the Shr\"odingier picture to the interaction picture.  Therefore, there is no such concept of ``propagator" for the purely classical laser beam.  However, it is reasonable to use the name ``propagator", $\hat{U}(t, t')$ also for the classical electromagnetic field,  since the time evolution factor is the same as in the quantum theory.  Then, the propagator of the classical laser beam is
\begin{eqnarray}
D^L_{C; C'}(t, z; t', z')=\frac{e^{-i \omega_L \{(t-t') - (z-z')/c')\} }}{1- r_{L}\left(e^{i \phi (\omega L/c)}\right)}\equiv e^{-i \omega_L \{(t-t') - (z-z')/c')\}} K^{\omega_L}_L (z, z'). \label{K2}
\end{eqnarray}
Here, we have ignored the damping effect of $\gamma'$, since it  should be cancelled by the strong pumping of the laser beam.  

 3) ``propagator" from the final interaction point to the outside photo-detector (PD):

This ``propagator" has an additional effect to the propagator, which is caused by passing through the mirror $M_2$, and by traveling from the final interaction point $z_f$ to the photo-detector located outside at a distance $z-L'/2$ from the mirror $M_2$.  

We denote the final interaction plane be $C_f$, then we have
\begin{eqnarray}
&&D^\omega_{PD; C_f}(t, z; t_f, z_f)=  \frac{\sqrt{1-r_2^2} \; e^{-i  \omega \{  (t-t_f)-(\ell/2-z_f)/c'-(z-L'/2)/c \} -\gamma'(\ell/2-z_f)}}{1- r_{\omega}e^{i \phi (\omega L/c)-\gamma'\ell} } \nonumber \\
&&\equiv e^{-i  \omega \{  (t-t_f)-(\ell/2-z_f)/c'-(z-L'/2)/c \} }K^{\omega}_{PD; C_f}(z, z_f). \label{K3}
\end{eqnarray}

The ``propagators'' include, other than the phase change, the additional factors $K$s, which describes the damping effect of photons in the cavity.  
\subsection{Vertex}
The real interaction of our problem occurs at the crystal planes separated by $a_{\mathrm{crys}}$, so that the ``vertex" is given in the vicinity of the crystal plane, giving
\begin{eqnarray}
 \langle m| \hat{H}_1(t)_S| m' \rangle =\int_{z-a_{\mathrm{crys}}/2}^{z+a_{\mathrm{crys}}/2} dz'\; \langle m |\hat{\mathcal{H}}_1(t, z')_S |m' \rangle,
\end{eqnarray}
where we have assumed the local interaction at $z$.  In the above expression we put the suffix $S$ as $\hat{\mathcal{H}}_1(t, z)_S$, to clarify the ``Shr\"odingier picture''. We assume as usual that the operators in the ``Shr\"odingier picture" and the ``interaction picture'' coincide at $t=0$. 

We will write the interaction Hamiltonian, after separating the dominant classical part of laser from the quantum part of photons, as follows:
\begin{eqnarray}
&&\langle m| \hat{H}_1(t)_S| m' \rangle=\int_{z-a_{\mathrm{crys}}/2}^{z+a_{\mathrm{crys}}/2} dz'\; \langle m|\hat{\mathcal{H}}_1(t, z')_S |m'\rangle \\
&&= \frac{1}{2} \varepsilon_0 \sum_{n \ge 2} n(n+1) \chi^{(n)} \int_{z-a_{\mathrm{crys}}/2}^{z+a_{\mathrm{crys}}/2} dz'\; E_L(t, z') \times \langle m| (\hat{E}_q(t=0, z'))^n |m' \rangle.
\end{eqnarray}

In the following, we study a simple NL case of having only $\chi^{(2)}$ and $\chi^{(4)}$, and hence the real interaction Hamiltonian takes the following form:
\begin{eqnarray}
\hat{\mathcal{H}}'_1(t, z)_{S}= \frac{1}{2} \varepsilon_0E_L(t, z) \times \left\{ 6 \chi^{(2)} (\hat{E}_q(0, z))^2 + 20 \chi^{(4)} (\hat{E}_q(0, z))^4 \right\}.
\end{eqnarray}

The real interaction occurs inside the non-linear crystal, where the light velocity is $c'$, so that the mode expansion used in writing the vertex function is\footnote{We do put $S$ for the classical laser beam as well as the quantum photons, in order to indicate that all the time dependencies have been absorbed into the classical and quantum propagators.}  

\begin{eqnarray}
\begin{cases} 
(\mathrm{classical~part~}): ~E_L(z)_S=  E_L \; e^{i\omega_L (z/c')}+E_L^{\dagger} \; e^{-i\omega_L (z/c')}, \\
(\mathrm{quantum~part}):~\hat{E}_q(z)_S= i \sum_{i=1, \; 2} \sqrt{\frac{\hbar \omega_i}{2L\varepsilon}}\left(\hat{a}_i e^{i\omega_i z/c'} - \hat{a}_i^{\dagger} e^{-i\omega_i z/c'} \right),
\end{cases}
\end{eqnarray}
where all the classical and quantum photons are assumed to propagate in the positive (clockwise) direction along the cavity, and the frequency is restricted to $\omega_1=\omega_L/4$ and $\omega_2=\omega_L/2$, by the resonance condition of the cavity.  

This is a good place to discuss the momentum conservation.  From the position integral near the vertex point, if the space near the vertex point is uniform ({\i.e.} the relevant wave length $\lambda$ of laser beam and photons satisfies $\lambda \ll a_{\mathrm{crys}}$, then the momentum conservation holds, and the sum of all the wave vectors coming in the vertex is equal to the sum of all the wave vectors coming out from the vertex. This property also holds in the usual old fashioned perturbation theory.  

As for the energy conservation, it does not hold in the old fashioned perturbation theory, since the transition amplitude becomes singular, if the energy conservation at each interaction point holds.  This problem is solved in this paper, by averaging the frequency of laser beam around the resonance point. (See the details in the next section.) 

Now, the possible interactions can be listed as follows:
\begin{eqnarray}
&&\chi^{(4)} : \begin{cases} 
(4)_4: \; \mathrm{Laser~emits}~4\gamma_1s~\mathrm{at~a~crystal~surface}, (\omega_L=4 \omega_1 \to 4\omega_1), \\
(4)_{-4}: \; 4 \gamma_1s~\mathrm{are~absorbed~by~Laser~at~the~surface}, (4 \omega_1 \to\omega_L=4\omega_1), \\
(4)_2: \; \mathrm{Laser}~+ 1\gamma_1~\mathrm{emitts}~2\gamma_2s+1\gamma_1 ~\mathrm{at~a~crystal~surface}, \\
(\omega_L+\omega_1 \to \omega_1+2 \omega_2 ),\\
(4)_{-2}: \; 2\gamma_2s+1\gamma_1~\mathrm{emit~Laser}+1\gamma_1~\mathrm{at~a~crystal~surface}, \\
(2 \omega_2+\omega_1 \to \omega_L + \omega_1).
\end{cases} \\
&&\chi^{(2)} : 
\begin{cases}
(2)_2; \;  \mathrm{Laser~emits}~ 2\gamma_2s~ \mathrm{at~a~crystal~surface}, (\omega_L=2 \omega_2 \to 2\omega_2), \\
(2)_{-2}: \; 2 \gamma_2s ~\mathrm{are~absorbed~by~Laser~at~the~surface}, (2\omega_2 \to 2\omega_2=\omega_L).
\end{cases} 
 \label{V}
\end{eqnarray}
where we denote the photons with energy $\omega_1$ and $\omega_2$ as $\gamma_1$ and $\gamma_2$, respectively.  The different types of interactions are labeled by $(4)_{\pm 4}, (4)_{\pm 2}$, and $(2)_{\pm 2}$ to make the individual processes clearer.  It is  noted that the other processes, such as $(\omega_L + 1 \gamma_1 \leftrightarrow 3 \gamma_1)$ does not occur, since the energy conservation is violated. 

Now, the various ``vertices" can be written down explicitly as follows: 
\begin{eqnarray}
\begin{cases}
(4)_4: \; ~\langle n_1, n_2 |\hat{H}_1(t) | n_1-4, n_2 \rangle =10 \varepsilon_0 \chi^{(4)}E_L e^{-i \omega_Lt} a_{\mathrm{crys}}\left( \frac{\hbar \omega_1 }{2L \varepsilon} \right)^2 \sqrt{(n_1-3)(n_1-2)(n_1-1)n_1}, \\
(4)_{-4}: \; \langle n_1, n_2 |\hat{H}_1(t) | n_1+4, n_2 \rangle =10 \varepsilon_0 \chi^{(4)}E_L^{\dagger}e^{i \omega_Lt} a_{\mathrm{crys}}\left( \frac{\hbar \omega_1}{2L \varepsilon} \right)^2 \sqrt{(n_1+1)(n_1+2)(n_1+3)(n_1+4)}, \\
(4)_2: \; ~\langle n_1, n_2 |\hat{H}_1(t) | n_1, n_2-2 \rangle =-10 \varepsilon_0 \chi^{(4)}E_L e^{-i \omega_Lt} a_{\mathrm{crys}}\left( \frac{\hbar \omega_1 }{2L \varepsilon} \right)\left( \frac{\hbar \omega_2 }{2L \varepsilon} \right) (2n_1+1) \sqrt{n_2(n_2-1)}, \\
(4)_{-2}: \; \langle n_1, n_2 |\hat{H}_1(t) | n_1, n_2+2 \rangle =-10 \varepsilon_0 \chi^{(4)}E_L^{\dagger}e^{i \omega_Lt} a_{\mathrm{crys}}\left( \frac{\hbar \omega_1 }{2L \varepsilon} \right)\left( \frac{\hbar \omega_2 }{2L \varepsilon} \right) (2n_1+1) \sqrt{(n_2+1)(n_2+2)}, \\
(2)_2:\;  ~\langle n_1, n_2 |\hat{H}_1(t) | n_1, n_2-2 \rangle =3 \varepsilon_0 \chi^{(2)}E_L e^{-i \omega_Lt} a_{\mathrm{crys}}\left( \frac{\hbar \omega_2}{2L\varepsilon} \right) \sqrt{(n_2-1)n_2}, \\
(2)_{-2}: \; \langle n_1, n_2 |\hat{H}_1(t) | n_1, n_2+2 \rangle =3 \varepsilon_0 \chi^{(2)}E_L^{\dagger}e^{i \omega_Lt} a_{\mathrm{crys}}\left( \frac{\hbar \omega_2}{2L \varepsilon} \right) \sqrt{(n_2+2)(n_2+1)}.
\end{cases}
\end{eqnarray}

\section{General Formula of Transition Amplitude or S-matrix in QO inside Cavity}
To obtain the general formula of transition amplitude or S-matrix, we first perform integrations over times $(t_1, t_2, t_3, \cdots)$ of the product of vertex operators, which are arranged according to the time ordering.  

Time dependence of the vertex is $E_L e^{-i\omega_L t}$ or $E_L^{\dagger} e^{i\omega_L t}$, and that of propagator $D(t-t')$ is $e^{\pm i\omega (t-t')}$, and hence the time integration is the same as in quantum mechanics.  Here, we write the vertex with ``energy'' eigen-states as $\langle E'_k |V(t_k) | E_k \rangle$, where $E_k$ and $E'_k$ are energies before and after the k-th interaction.  Here we do not distinguish the ``energy'' and the ``angular frequency''.

Then, up to position dependent extra factors, $K$s, the transition amplitude $S(t, -\infty)$, defined by $\psi(t)_f=S(t, -\infty) \; \psi(-\infty)_i$, reads perturbatively, 
\begin{eqnarray}
&&S(t, -\infty)\vert_{\mathrm{up\;to}\;Ks}=1 + \sum_{N=1}^{\infty} \left(\frac{1}{i \hbar} \right)^N \int_{-\infty}^{t} dt_1 \int_{-\infty}^{t_1} dt_2 \cdots \int_{-\infty}^{t_{N-1}} dt_N \;   e^{-i(\omega^L_1+E_1-E_f+i \epsilon)t_1} \langle E_f |\hat{H}_1 | E_1 \rangle  \nonumber \\ 
&&~~~~~~~~~~~~~~~~~~~~~~ \times e^{-i(\omega^L_2 +E_2-E'_2+i \epsilon)t_2} 
 \langle E'_2 | \hat{H}_1 | E_2 \rangle \times  \cdots  \times e^{-i(\omega^L_N+E_i-E'_N+i \epsilon)t_N} \langle E'_N |\hat{H}_1 | E_i \rangle \nonumber \\
&&=1 + \sum_{N=1}^{\infty}  e^{-i(\omega^L_1 + \cdots + \omega^L_N+E_i-E_f)t} \prod_{k=1}^{N} \frac{\langle E'_k |\hat{H}_1 | E_k \rangle/\hbar}{(\omega^L_k+ \cdots + \omega^L_N) +E_i-E'_k+i\epsilon},~~~~\label{old fashioned perturbation}
\end{eqnarray}
where $E_i=E_N$ and $E_f=E'_1$ are initial and final ``energies'' of photons, respectively.  The ``energies'' before and after each interaction can be expressed by using the number representation as $\langle (n_k)_1, (n_k)_2 | \hat{H}_1(t_k)|(n_k)_1-(s_k)_1, (n_k)_2-(s_k)_2 \rangle$, where the suffix 1 and 2 indicate the particle numbers $(n_k)_{1,2}$ and their difference $(s_k)_{1,2}$ be for the two kinds of photons, $\gamma_1$ and $\gamma_2$, respectively.  Then, we have
\begin{eqnarray}
E'_k=E_i+ (s_{k}+s_{k+1} + \cdots + s_{N})_1 \; \omega_1+(s_{k}+s_{k+1} + \cdots +s_{N})_2 \; \omega_2,
\end{eqnarray}
where $\omega_1$ and $\omega_2$ are frequencies of photon $\gamma_1$ and $\gamma_2$, respectively. 

The formula obtained as Eq.(\ref{old fashioned perturbation}) is nothing but the formula of old fashioned perturbation theory in QM.  When we apply it to our QO, the phase matching condition tells that the energy and momentum conservation hold at each vertex, which leads to the vanishing of all the denominators in the formula, giving a singular ({\it i. e.} extremely dominant) behavior for the scattering amplitude.  This singularity is called the ``mass singularity''; it happens when the involved particles are massless like photons, and a particle decays into particles moving in the same direction to the parent particle. (See the reference \cite{Kinoshita-Lee-Nauenberg})

A way to tame this singularity is to average the transition amplitude over the laser frequency $\omega^L_k$ at each interaction point $k$.  This averaging is a kind of tuning of the laser beam at the resonance point.

A candidate of the weight for this averaging is a Gaussian distribution, $W(|\omega_L|)$, with a center $\omega_R$ and a width $\delta$,\footnote{Corresponding to absorption and emission of laser light, we consider $\omega^L$ as positive and negative values, respectively.}
\begin{eqnarray}
W(|\omega^L_k|)=\frac{1}{\sqrt{2\pi}\;\delta} \; e^{-(|\omega^L_k|-\omega_R)^2/2\delta^2}.
\end{eqnarray}
This averaging implies that the laser photon has a definite wave vector $k^L=\pm \omega_R/c'$ at the interaction plane of crystal, but its energy is not exactly $\pm \omega_R$, but is distributed as\footnote{At first sight, this averaging seems to violate the ``massless property" of photon, but is a standard way to tame the mass singularity.  Physically, this $\delta$ can be the ambiguity to detect photon energy, which is inevitable.  In QO, it can be the ``ambiguity of tuning the laser frequency at the resonance point''.  It can also be that the laser beam is not a plane wave but the wave packet.}
\begin{eqnarray}
\omega_R-\delta < |\omega^L_k| < \omega_R+\delta.
\end{eqnarray}

Then, the averaging over the laser frequency becomes
\begin{eqnarray}
\langle S(t, -\infty)\vert_{\mathrm{up\;to}\;Ks} \rangle_{\omega_L}  = 1+ \sum_{N=1}^{\infty}\prod_{k=1}^N \int_{-\infty}^{\infty} d \; \omega^L_k \; W(|\omega^L_k|) S(t, -\infty; \{\omega^L_k\}) \vert_{\mathrm{up\;to}\;Ks}.
\end{eqnarray} 

Since the weight function vanishes at infinitely large frequency, 
\begin{eqnarray}
W(|\omega^L_k|) \to 0,~~\mathrm{for}~~ |\omega^L_k| \to \infty.
\end{eqnarray}
we can perform the contour integration clock-wisely by the new variables with tilde, defined by
\begin{eqnarray}
\tilde{\omega}^L_1=\omega^L_1+ \cdots \omega^L_N, \; \tilde{\omega}^L_2=\omega^L_2+ \cdots \omega^L_N, \cdots, \; \tilde{\omega}^L_N=\omega^L_N,
\end{eqnarray}
and pickup the pole residues.  The choice of pole residues recovers the energy conservation at each vertex, $\omega^L_k=E'_k-E_k$, but $\omega^L_k$ is not definitely fixed, but is distributed with a sharp peak around $\pm \omega_R$.   Then, we arrive at 
\begin{eqnarray}
\langle S(t, -\infty)\vert_{\mathrm{up\;to}\;Ks} \rangle_{\omega_L}=1 + \sum_{N=1}^{\infty} \; \prod_{k=1}^{N} \; (-2\pi i/\hbar) W(|E'_k-E_k|) \langle E'_k |\hat{H}_1 | E_k \rangle,
\end{eqnarray}
where 
\begin{eqnarray}
W(|E'_k-E_k|)=\frac{1}{\sqrt{2\pi}\; \delta} \; e^{-(|E'_k-E_k|-\omega_R)^2/2\delta^2},  
\end{eqnarray}
where the pre-factor $e^{-i(\omega^L_1 + \cdots + \omega^L_N+E_i-E_f)t}=1$ holds, and we can take $t \to \infty$ without problem.\footnote{Here, we have to note that in doing the contour integral in the complex plane of $\tilde{\omega}^L_k$, no other poles appear, even if the factor $K$s are included, since the denominator of the factor $K_L^{\omega_L}(z, z')$ is $1- r_L e^{i \phi(\omega_L L/c)}$,
and it can not be zero, unless the mirror perfectly reflects the laser beam ($r_L < 1$).}  If the resonance peak is very sharp, we are allowed to approximate the Gaussian distribution at its peak value,
\begin{eqnarray}
W(\omega^L_k) \approx \frac{1}{\sqrt{2\pi}\;\delta}, ~\mathrm{with~the~condition}~~E'_k=E_k \pm \omega_R.
\end{eqnarray}

Now, we are ready to include the factors $K(z, z')$s, being ignored in the above, and to obtain the final formula of transition amplitude in QO.

The formula for the transition amplitude $\langle S(t, -\infty) \rangle_{\omega_L}$ for $t=\infty$, or the S-matrix $\langle S(\infty, -\infty)  \rangle_{\omega_L}$ in QO is obtained as follows:
\begin{eqnarray}
&&\langle E_f|\hat{S}|E_i\rangle=1+ \sum_{N=1}^{\infty} \left(\frac{-2\pi i}{\hbar \sqrt{2 \pi} \; \delta}\right)^N \prod_{k=1}^N  \int_{-L/2}^{L/2} dz_1 dz_2 \cdots dz_k  \; K(z, z_1)_{PD;C_1} \langle E_f |\hat{H}_1(0) | E_1 \rangle  \nonumber \\
&&~~~~~~~~~~~~~~~~~~~\times K_{C_1;C_2}(z_1, z_2) K^L(z_1, z_2) \langle E'_2 |\hat{H}_1(0)  | E_2 \rangle K_{C_2;C_3}(z_2, z_3) K^L(z_2, z_3) \times \cdots \nonumber \\
&&~~~~~~~~~~~~~~~~~~\times K_{C_{N-1};C_{N}}(z_{N-1}, z_N) K^L(z_{N-1}, z_N)\langle E'_N |\hat{H}_1(0)  | E_i \rangle K_{C_N;C_0}(z_N, z_0),
\end{eqnarray}
where the energy difference can be taken as $E'_k-E_k=\pm \omega_R$ at each vertex, and $\delta/2$ is the width of laser frequency distribution around the resonance point $\omega_R$.  The factors $K$s are given in Eqs.(\ref{K1})-(\ref{K3}), and the vertices $\langle E' |\hat{H}_1(0)  | E \rangle$ are in Eq.(\ref{V}), being categorized into six types, $(4)_{\pm 4}, (4)_{\pm 2}$ and $(2)_{\pm 2}$.

Before ending this section, let us summarize the expressions for $K$s and vertices:

The damping factors $Ks$ are
\begin{eqnarray}
K=\begin{cases}
K^{\omega}_{C;C'}(z, z') =\frac{e^{-\gamma'|z-z'|}}{1- \left(r_{\omega} e^{i \phi (\omega L/c)-\gamma'\ell}\right) }, \\
K^{\omega}_L (z, z')=\frac{1}{1- r_{L}\left(e^{i \phi (\omega L/c)}\right)}, \label{K2'}\\
K^{\omega}_{PD; C_f}(z, z_f)=  \frac{\sqrt{1-r_2^2} \; e^{-\gamma'(\ell/2-z_f)}}{1- r_{\omega}e^{i\phi (\omega L/c)-\gamma'\ell} }, \label{K3'}
\end{cases}
\end{eqnarray}
where $r_{\omega, \; \omega_L}=(r_1r_2r_3r_4)_{\omega, \; \omega_L}$,  and $\phi(\omega L/c)=\frac{\omega}{c}(L-\ell) + \frac{\omega}{c'} \ell$ ($\ell$: the length of the non-linear crystal, $L$: the circumference of the ring resonator).  

The vertices $V$s are 
\begin{eqnarray}
\begin{cases}
(4)_4: \; ~\langle n_1, n_2 |\hat{H}_1(0) | n_1-4, n_2 \rangle =10 \varepsilon_0 \chi^{(4)}\left(\frac{\hbar\omega_R}{8L \varepsilon}\right)^2 a_{\mathrm{crys}}\;E_L \sqrt{(n_1-3)(n_1-2)(n_1-1)n_1}, \\
(4)_{-4}: \; \langle n_1, n_2 |\hat{H}_1(0) | n_1+4, n_2 \rangle =10 \varepsilon_0 \chi^{(4)} \left(\frac{\hbar\omega_R}{8L \varepsilon}\right)^2 a_{\mathrm{crys}}\; E_L^{\dagger} \sqrt{(n_1+1)(n_1+2)(n_1+3)(n_1+4)}, \\
(4)_2: \; ~\langle n_1, n_2 |\hat{H}_1(0) | n_1, n_2-2 \rangle =-10 \varepsilon_0 \chi^{(4)} \left(\frac{\hbar\omega_R}{4L \varepsilon}\right)^2 a_{\mathrm{crys}}\;E_L (2n_1+1) \sqrt{n_2(n_2-1)}, \\
(4)_{-2}: \; \langle n_1, n_2 |\hat{H}_1(0) | n_1, n_2+2 \rangle =-10 \varepsilon_0 \chi^{(4)}\left(\frac{\hbar\omega_R}{4L \varepsilon}\right)^2 a_{\mathrm{crys}}\; E_L^{\dagger} (2n_1+1)\sqrt{(n_2+2)(n_2+1)} , \\
(2)_2:\;  ~\langle n_1, n_2 |\hat{H}_1(0) | n_1, n_2-2 \rangle =3 \varepsilon_0 \chi^{(2)} \left(\frac{\hbar\omega_R}{4L \varepsilon}\right) a_{\mathrm{crys}}\;E_L  \sqrt{(n_2-1)n_2}, \\
(2)_{-2}: \; \langle n_1, n_2 |\hat{H}_1(0) | n_1, n_2+2 \rangle =3 \varepsilon_0 \chi^{(2)} \left(\frac{\hbar\omega_R}{4L \varepsilon}\right) a_{\mathrm{crys}}\; E_L^{\dagger}  \sqrt{(n_2+2)(n_2+1)}.
\end{cases}
\end{eqnarray}

Now, the general formula has been obtained, which is useful in quantum optics (QO) to estimate the higher-order effects inside the ring resonator with a non-linear material contained.

\section{Conclusion and Discussion}
We intend in this paper to derive a general formula to be used to estimate the higher-order effects in QO, especially in a ring resonator with a non-linear material contained.  The obtained formula is expressed Feynman graphically using propagators and vertices, in which the propagators include the effects of damping in the non-linear material as well as the reflection and transmission by mirrors.  See Appendices A, B and C for the relation of our method to the other works.

We will discuss the possible observables being calculable in principle in the above-mentioned QO.

(1) Average number of produced photons:  

We consider a simple case of NL-material with non-linear susceptibility, $\chi^{(2)}$ and $\chi^{(4)}$.  For the laser beam with a resonance energy $\omega^L$, two photons $\gamma_1$ and $\gamma_2$ are relevant, having energies $\omega^L/4$ and $\omega^L/2$, respectively.  We can estimate the averaged number of photons $\langle n_1 \rangle$ and $\langle n_2\rangle $ for $\gamma_1$ and $\gamma_2$, as well as the correlations, $\langle (n_1)^{m_1}(n_2)^{m_2}  \rangle, ~ (m_1, m_2)=1, 2, \cdots$.  To do this we have to use the numerical simulations, and to compare the result so obtained with a theoretical study of the ``generalized squeezing'' in \cite{Katagiri}.  Also refer to Appendix D.

(2) Husimi function:

Averaged number of the produced photons is directly connected to the Husimi function $H(\alpha)$ \cite{Husimi}. (Please refer to Wigner function \cite{Wigner}.) It is a quantum version of the distribution function and is called ``Q function" $Q(\alpha)$ in QO:
\begin{eqnarray}
H(t, \alpha) \equiv Q(t, \alpha)= \langle \alpha | \hat{\rho}(t) | \alpha \rangle,
\end{eqnarray}
where $\hat{\rho}(t)$ is a density matrix at time $t$, and $| \alpha \rangle$ is a coherent state with a complex eigen-value $\alpha$ for the annihilation operator $\hat{a}$.

In our case, we have two sets of annihilation operators, so that the Husimi function can be written at $t=\infty$ in terns of $S$-matrices as follows:
\begin{eqnarray}
&&H(\alpha_1, \alpha_2) =\langle \alpha_1, \alpha_2| \hat{S} |0 \rangle \langle \alpha_1, \alpha_2 | \hat{S} |0 \rangle^{*}= e^{-(|\alpha_1|^2+|\alpha|_2^2)} \nonumber \\
&&\times \sum_{\{\bar{n}_{1,2}, \bar{m}_{1,2}\}=1}^{\infty} \frac{(\alpha_1^{*})^{2\bar{n}_1} (\alpha_1)^{2\bar{m}_1}(\alpha_2^{*})^{2\bar{n}_2} (\alpha_2)^{2\bar{m}_2}}{(2\bar{n}_1)! (2\bar{m}_1)!(2\bar{n}_2)! (2\bar{m}_2)}  \times \langle 2\bar{n}_1, 2\bar{n}_2 |\hat{S} |0 \rangle \langle 2\bar{m}_1, 2\bar{m}_2| \hat{S} |0 \rangle^{*},  \label{Husimi function}
\end{eqnarray}

(3) Observable to measure beyond the squeezing:

If we introduce a phase shifted wave, 
\begin{eqnarray}
&&E^{\theta}(t, z)=  e^{-i \theta} E \; e^{-i\omega_L (t-z/c(z))} +  e^{i \theta} E^{\dagger}\; e^{i\omega_L (t-z/c(z))}, 
 \end{eqnarray}
then, it can accommodate the vector potential $A(t, z) \propto E^{\theta=\pi/2}(t, z)$ and the electric field $E(t, z) = E^{\theta=0}(t, z)$, which are canonical conjugate with each other.  Therefore, the correlation functions of the phase sifted waves described the squeezing.   

In our case, there exist various two-point and three-point correlation functions, since the decay of $\gamma_2 \to \gamma_1 + \gamma_1$ is possible by $\omega_2=2\omega_1$,   
\begin{eqnarray}
\begin{cases}
\langle \Delta E^{\theta}_1(t) \Delta E^{\theta}_1(t') \rangle_{PD}=C^{\theta}_{11}(t, t'), ~\langle  \Delta E^{\theta}_2(t) \Delta E^{\theta}_2(t') \rangle_{PD}=C^{\theta}_{22}(t, t'),~\mathrm{for}~(t > t'), \\
\langle  T [\Delta E^{\theta}_1(t_1) E^{\theta}_1(t'_1)\Delta E^{\theta} _2(t_2) ] \rangle_{PD}=C^{\theta}_{T[1, 1', 2]}(T[t_1, t_1', t_2]),
\end{cases}
\end{eqnarray}
the last three-point correlation function gives, by specifying the ordering of time, as
\begin{eqnarray}
\begin{cases}
\langle  \Delta E^{\theta}_1(t_1) E^{\theta}_1(t'_1)\Delta E^{\theta} _2(t_2) \rangle_{PD}=C^{\theta}_{112}(t_1, t_1', t_2)~\mathrm{for}~(t_1 > t'_1>t_2), \\
\langle  \Delta E^{\theta}_1(t_1)\Delta E^{\theta} _2(t_2)  E^{\theta}_1(t'_1)\rangle_{PD}=C^{\theta}_{121}(t_1, t_2, t_1')~\mathrm{for}~(t_1 >t_2 > t'_1), \\
\langle   \Delta E^{\theta} _2(t_2) \Delta E^{\theta}_1(t_1) \Delta E^{\theta}_1(t'_1) \rangle_{PD}=C^{\theta}_{211}(t_2, t_1, t_1')~\mathrm{for}~(t_2 > t_1>t'_1).
\end{cases}
\end{eqnarray} 

Corresponding to the above correlation functions, we can introduce various spectra:
\begin{eqnarray}
\begin{cases}
S_{11}(\nu, \theta)=\int_{0}^{\infty} d\tau~C^{\theta}_{11}(t, t'=t +\tau) e^{-i\nu \tau}, \\
S_{22}(\nu, \theta)=\int_{0}^{\infty} d\tau~C^{\theta}_{22}(t, t'=t +\tau) e^{-i\nu \tau}, \\
S_{112}(\mu, \nu, \theta)=\int_{0}^{\infty} d\tau \int_{0}^{\infty} d\sigma~C^{\theta}_{112}(t_1, t'_1=t_1+ \tau, t_2=t'_1 + \sigma ) e^{-i(\mu \tau+\nu \sigma)},\\
S_{121}(\mu, \nu, \theta)=\int_{0}^{\infty} d\tau \int_{0}^{\infty} d\sigma~C^{\theta}_{121}(t_1, t_2=t_1+\tau, t_1'=t_1+\sigma) e^{-i(\mu \tau+\nu \sigma)}, \\
S_{211}(\mu, \nu, \theta)=\int_{0}^{\infty} d\tau \int_{0}^{\infty} d\sigma~C^{\theta}_{211}(t_2, t_1=t_2+\tau, t_1'=t_1+\sigma) e^{-i(\mu \tau+\nu \sigma)}
\end{cases}
\end{eqnarray}

The second spectrum $S_{22}(\nu, \theta)$ was measured by L-A Wu {\it et al.}\cite{Wu} using the OPO which contains the non-linear crystal with non-vanishing $\chi^{(2)}$. This measurement shows the squeezing is realized in OPO. The spectrum $S_{22}(\nu, \theta)$ represents the down-conversion of the laser photon to two degenerate photons $(\omega_L  \to 2\omega_2)$. 

Therefore, it is interesting to examine the new spectra $S_{112}(\mu, \nu, \theta)$, $S_{121}(\mu, \nu, \theta)$ and $S_{211}(\mu, \nu, \theta)$, in addition to $S_{11}(\nu, \theta), S_{22}(\nu, \theta)$, in case of NL-crystal with non-vanishing $\chi^{(2)}$ and $\chi^{(4)}$.  

We hope to study these issues in the next paper of ours \cite{our next paper}.


\section*{Appendix A: -Input-output theory and observation of squeezed state in optical cavity-}

In this Appendix, we review Input-Output Theory and squeezed state observation in optical cavity based on Corrett and Gardiner \cite{key-2, key-4}, and compare it to our paper.  Even if this Input-Output theory is popular in the nonlinear optics, it is not a rigorous theory, but is rather a phenomenological one.  We will explain this, by following the theory. 

The input electric field enters from outside into the optical cavity,
interacts with it, and emits the output photon, which is observed by the photo-detector (PD).

Suppose that a single mode $\Omega$ of photon remains in the optical cavity,
where the creation and annihilation operators of the photon are denoted by $\hat{a}^{\dagger},\hat{a}$. The cavity (Fabry-P\'erot type) is assumed to have a perfect reflection on one side and a reflection mirror with transmittance on the other side.

In this situation, the Hamiltonian is divided into three parts,

\begin{equation}
\hat{H}=\hat{H}_{Cavity}+\hat{H}_{External}+\hat{H}_{Int}, \label{eq:InputOutputHamiltonian}
\end{equation}
where, $\hat{H}_{Cavity}$ is the Hamiltonian of the optical cavity, $\hat{H}_{External}$ is that outside the cavity, and $\hat{H}_{Int}$
is the interaction Hamiltonian between the inside and the outside of the cavity.

We can give $\text{\ensuremath{\hat{H}}}_{Int}$ as follows:

\begin{equation}
\hat{H}_{Int}=\hat{\Gamma}\hat{a}^{\dagger}(t)+\hat{\Gamma}^{\dagger}\hat{a}(t),
\end{equation}

\begin{equation}
\hat{\Gamma}=i\hbar\int_{-\infty}^{\infty}d\omega\kappa(\omega)b_{\omega}(t),
\end{equation}
where, $\hat{b}^{\dagger}_{\omega}(t),\ \hat{b}_{\omega}(t)$ are the creation-annihilation operators of photons in the external electric field.  The form of interaction shows this is a toy model, describing a transition of photon, $\gamma_{\Omega} \to \gamma_{\omega}$, admitting the loss of energy ({\it e.g.} a phenomenological inelastic interaction, violating energy conservation). 

Then, Heisenberg equations are

\begin{equation}
\begin{cases}
~\dot{\hat{a}}(t)=-\frac{i}{\hbar}[\hat{a}(t),\hat{H}_{Cavity}]-\int_{-\infty}^{\infty}d\omega \; \kappa(\omega)  \hat{b}_{\omega}(t), \\\label{eq:heisenbergForA}
~\dot{\hat{b}}_{\omega}(t)=-\frac{i}{\hbar}[\hat{b}_{\omega}(t),\hat{H}_{External}]+\kappa(\omega)\hat{a}(t), \end{cases}
\end{equation}
where, we suppose $\hat{H}_{External}$ is free, then 
\begin{equation}
\dot{\hat{b}}_{\omega}(t)=-i\omega\hat{b}_{\omega}(t)+\kappa(\omega)\hat{a}(t).
\end{equation}

We are considering the Heisenberg picture, and should be careful about the time dependency, so that we have written the $t$-dependence explicitly at delicate places. We take the initial condition $b_{\omega, 0}=b_{\omega}(t_0)$ at $t=t_{0}$ and final condition $b_{\omega, 1}=b_{\omega}(t_1)$ at $t=t_{1}$. 

The solution to the above equations is
\begin{equation}
\begin{cases}
~\hat{b}_{\omega}(t)=e^{-i\omega\left(t-t_{0}\right)}\hat{b}_{ \omega, 0}+\kappa\left(\omega\right)\int_{t_{0}}^{t}e^{-i\omega(t-t')}\hat{a}\left(t'\right)dt'\ (t_{0}<t), \\
~\hat{b}_{\omega}(t)=e^{-i\omega\left(t-t_{1}\right)}\hat{b}_{\omega, 1}-\kappa\left(\omega\right)\int_{t}^{t_{1}}e^{-i\omega\left(t-t'\right)}\hat{a}(t')dt'\ \left(t<t_{1}\right). \end{cases}
\end{equation}

Unfortunately, this equation is not rigorous, but is valid only in the first order approximation.  The reason can be understood from a well-known fact that if the Hamiltonian controlling $\partial_t \hat{b}_{\omega}$ is $t$-dependent as $\kappa(\omega) \hat{a}(t)$, the rigorous solution is expressed in terms of T-ordered product of $e^{\int_{t_0}^t dt' \kappa(\omega) \hat{a}(t')}$, or anti-T-ordered product of $e^{\int_{t_1}^t dt' \kappa(\omega) \hat{a}(t')}$,\color{black}

We suppose $\kappa(\omega)^{2}=\gamma/2\pi$, then 
(\ref{eq:heisenbergForA}) give two expressions, 
\begin{equation}
\begin{cases}
~\dot{\hat{a}}=-\frac{i}{\hbar}[\hat{a}(t),\hat{H}_{Cavity}]+\sqrt{\gamma}\hat{a}_{IN}(t)-\frac{\gamma}{2}\hat{a}(t), \\
~\dot{\hat{a}}=-\frac{i}{\hbar}[\hat{a}(t),\hat{H}_{Cavity}]-\sqrt{\gamma}\hat{a}_{OUT}(t) + \frac{\gamma}{2}\hat{a}(t), \end{cases}
\end{equation}
where
\begin{equation}
\begin{cases}
~\hat{a}_{IN}(t)\equiv-\frac{1}{\sqrt{2\pi}}\int_{-\infty}^{\infty}d\omega e^{-i\omega(t-t_{0})}b_{0}(\omega) \\
~\hat{a}_{OUT}(t)\equiv\frac{1}{\sqrt{2\pi}}\int_{-\infty}^{\infty}d\omega e^{-i\omega(t-t_{1})}b_{1}(\omega). \end{cases}
\end{equation}
Here, $\hat{a}_{IN}$ and $\hat{a}_{OUT}$ are called the Input and
Output fields, respectively and have the relationship:

\begin{equation}
\hat{a}_{OUT}(t)+\hat{a}_{IN}(t)=\sqrt{\gamma}\hat{a}(t).\label{eq:InputOutput}
\end{equation}

To derive these equations we have to suppose that $2\pi \kappa(\omega)^2=\gamma$ does not depends on $\omega$.
Therefore, the above equations on which the input-output theory is based, are phenomenological equations, derived with three assumptions, the phenomenological interaction, the first order perturbation theory and the $\omega$ independence of $\kappa$. 

From these, the two-time correlation functions of the input and output
fields can be obtained from the correlation functions of
the photons inside the optical cavity,
\begin{eqnarray}
\begin{cases}
~\langle\hat{a}_{OUT}^{\dagger}(t),\hat{a}_{OUT}(t')\rangle=\gamma\langle\hat{a}^{\dagger}(t),\hat{a}(t')\rangle, \\
~\langle\hat{a}_{OUT}(t),\hat{a}_{OUT}(t')\rangle=\gamma\langle\hat{a}(t),\hat{a}(t')\rangle+\gamma\theta(t'-t)\langle[\hat{a}(t'),\hat{a}(t)]\rangle,
\end{cases}
\end{eqnarray}
where $\langle\hat{U},\hat{V}\rangle\equiv\langle\hat{U}\hat{V}\rangle-\langle\hat{U}\rangle\langle\hat{V}\rangle$.
 
The phase shifted operator $\hat{X}_{\theta}$ is convenient to discuss the uncertainty relation between gauge field and its canonical conjugate,  electric field:
 \begin{equation}
\hat{X}_{\theta}(t)=e^{-i \theta} \hat{a}(t)+ e^{i \theta}\hat{a}^{\dagger}(t),
\end{equation}

The spectrum of the output photons is defined as a Fourier transform of the two-point correlation between the wave shifted fields, 
\begin{eqnarray}
&&S_{\theta}^{OUT}(\omega)-1=:S_{\theta}^{OUT}\left(\omega\right) :\equiv\int dt\langle:\hat{X}_{\theta}^{OUT}(t),\hat{X}_{\theta}^{OUT}(0):\rangle e^{-i\omega t} \\
&&\hspace{4.7cm}=\gamma\int dt \; T\langle:X_{\theta}(t),X_{\theta}(0):\rangle e^{-i\omega t}.
\end{eqnarray}
where $:\hat{A}\hat{B}:$ represents the normal order product and
$T$ is the time order product.

Using this spectrum of the input-output theory, L.-A. Wu {\it et al.} \cite{Wu} shows the squeezing by OPO.

\subsection{Relation of our paper to this Appendix A}
The final result of the input-output theory \cite{key-2, key-4} is very simple, that is, the electro-magnetic wave outside the cavity is damped from that inside the cavity by a damping factor $\gamma=\gamma_{\mathrm{(input-output)}}$.  This damping is caused by the penetration of the wave through the last mirror.  If so, it is enough to take into account the damping effect of the wave by the mirror, phenomenologically. As was discussed in the above, the input-output theory is based on a number of assumptions, and is not a rigorous theory.  We have a lot of sources of dispersion and damping; one is the scattering of the wave by the crystal surfaces of the non-linear optical material, and the others are the reflection by the mirrors and the penetration through them.  In this paper we consider these dispersive and damping effects phenomenologically, by using the measured values of reflection coefficient $r_i$ and transmission coefficient $t_i=\sqrt{1-r_i^2}$ for the $i$-th mirror.  The $\gamma_{\mathrm{(input-output)}}$ in the input-output theory for the final mirror is equal to the transmission coefficient $t_{\mathrm{final-mirror}}$ in our notation. 

In our paper we separate this part of dispersion and damping as the ``quasi-free Hamiltonian'', and concentrate on the ``real interaction Hamiltonian'' of photons with the non-linear material, so that we can cope with the higher order calculations by many times round trips of photons inside the cavity. 

\section*{Appendix B: -Fokker-Planck equation-}

In quantum mechanics with loss, a standard method is to consider that the system is surrounded by a heat bath (or reservoir), so that the energy of the system can dissipate to the heat bath. In this situation, the master equation (q-number equation) satisfied by the density matrix is used, which is transformed by a coherent representation to the Fokker-Planck equation (c-number equation). The Fokker-Planck equation can be rewritten as the Langevin equation. \cite{Fokker-Planck, gravity analog}

The Hamiltonian (\ref{eq:InputOutputHamiltonian}) in such a system
is divided into parts,

\begin{equation}
\hat{H}=\hat{H}_{Cavity}+\hat{H}_{External}+\hat{H}_{Int}\equiv\hat{H}_{Cavity,0}+\hat{H}_{Cavity,I}+\hat{H}_{External}+\hat{H}_{Int}\equiv\hat{H}_{0}+\hat{H}_{I}
\end{equation}
\begin{equation}
\hat{H}_{I}=\hat{H}_{Cavity,I}+\hat{H}_{Int},
\end{equation}
where $\hat{H}_{Cavity,0}$ is a free part of $\hat{H}_{Cavity}$.  The
state can be written as the tensor product of the output photon
state of $\mathcal{H}_{b}$, and the photon state in the resonator of $\mathcal{H}_{a}$, so that the total density operator is the tensor product of the density operator for the heat bath $b$ and that for the system ($a$), $\rho_{Total}=\hat{\rho}_{b}\otimes\hat{\rho}$.

Given the density matrix in the interaction picture $\rho_{Total}(t)$,
the equation of motion is
\begin{equation}
\frac{d\hat{\rho}_{Total}(t)}{dt}=-\frac{i}{\hbar}[\hat{H}_{I}(t),\hat{\rho}_{Total}(t)].
\end{equation}

This master equation can be solved as a formal expansion as 
\begin{equation}
\begin{aligned}\hat{\rho}_{Total}(t)= & \hat{\rho}_{Total}(0)+\sum_{n=1}^{\infty}\left(-\frac{i}{\hbar}\right)^{n}\int_{0}^{t}dt_{1}\int_{0}^{t_{1}}dt_{2}\dots\int_{0}^{t_{n-1}}dt_{n}\\
 & [\hat{H}_{Int}(t_{1}),[\hat{H}_{Int}(t_{2}),\dots,[\hat{H}_{I}(t_{n}),\hat{\rho}_{Total}(0)]\dots]]\\
\equiv & \left(1+\sum_{n=1}^{\infty}\hat{U}_{n}(t)\right)\left(\hat{\rho}(0)\right)\equiv\hat{U}(\hat{\rho}(0)).
\end{aligned}
\end{equation}

We do not observe the heat bath described by $\mathcal{H}_b$, so that we trace off the contribution from the heat bath, giving the density operator of the system, 
\begin{equation}
\hat{\rho}(t)=\mathrm{Tr}_{\mathcal{H}_{b}} \; \hat{\rho}_{Total}(t),
\end{equation}
where we assume
\begin{equation}
\mathrm{Tr}_{\mathcal{H}_{b}}\left(\hat{H}_{I}(t)\hat{\rho}_{b}\right)=0.
\end{equation}

Unfortunately in the actual application, we have to restrict up to $n=2$ for $U_{n}$, and discard the higher order contributions.  Then, the master equation for $\hat{\rho}$ becomes
\begin{equation}
\frac{d}{dt}\hat{\rho}(t)=-\frac{1}{\hbar^{2}}\int_{0}^{t}dt_{1}\mathrm{Tr}_{\mathcal{H}_{b}}\left[\hat{H}_{I}(t),[\hat{H}_{I}(t_{1}),\hat{\rho}_{b}\otimes\hat{\rho}(0)]\right].
\end{equation}

In the following, we will omit $\hat{H}_{Cavity,I}$.  Since $\text{\ensuremath{\hat{H}}}_{Int}$ (of a toy model) is given as
\begin{eqnarray}
\hat{H}_{Int}=\hat{\Gamma}\hat{a}^{\dagger}+\hat{\Gamma}^{\dagger}\hat{a}, ~~\hat{\Gamma}=i\hbar\int_{-\infty}^{\infty}d\omega \; \kappa(\omega)b_{\omega},
\end{eqnarray}
the master equation becomes
\begin{equation}
\begin{aligned}\frac{d}{dt}\hat{\rho}(t)= & -\frac{1}{\hbar^{2}}\left[I_{1}(t)\hat{a}^{\dagger}\hat{a}^{\dagger}\hat{\rho}(0)+I_{3}(t)\hat{a}^{\dagger}\hat{a}\hat{\rho}(0)-\left(I_{1}(t)\hat{a}^{\dagger}\hat{\rho}(0)\hat{a}^{\dagger}+I_{4}^{*}(t)\hat{a}^{\dagger}\hat{\rho}(0)\hat{a}\right)\right.\\
 & +I_{4}(t)\hat{\rho}_{b}\hat{a}\hat{a}^{\dagger}\hat{\rho}(0)+I_{2}(t)\hat{\rho}_{b}\hat{a}\hat{a}\hat{\rho}(0)-\left(I_{3}^{*}\hat{a}\hat{\rho}(0)\hat{a}^{\dagger}+I_{2}\hat{a}\hat{\rho}(0)\hat{a}\right)\\
 & -I_{1}(t)\hat{a}^{\dagger}\hat{\rho}(0)\hat{a}^{\dagger}-I_{3}(t)\hat{a}\hat{\rho}(0)\hat{a}^{\dagger}+\left(I_{1}(t)\hat{\rho}(0)\hat{a}^{\dagger}\hat{a}^{\dagger}+I_{4}^{*}(t)\hat{\rho}(0)\hat{a}\hat{a}^{\dagger}\right)\\
 & \left.-I_{4}(t)\hat{a}^{\dagger}\hat{a}\hat{\rho}(0)-I_{2}(t)\hat{a}\hat{\rho}(0)\hat{a}+\left(I_{3}^{*}(t)\hat{\rho}(0)\hat{a}^{\dagger}\hat{a}+I_{2}(t)\hat{\rho}(0)\hat{a}\hat{a}\right)\right],
\end{aligned}
\end{equation}
where
\begin{equation}
\begin{cases}
~I_{1}=\int_{0}^{t}dt_{1}\langle\hat{\Gamma}(t)\hat{\Gamma}(t_{1})\rangle e^{i\Omega(t+t_{1})}, \\
~I_{2}=\int_{0}^{t}dt_{1}\langle\hat{\Gamma}^{\dagger}(t)\hat{\Gamma}^{\dagger}(t_{1})\rangle e^{-i\Omega(t+t_{1})}, \\
~I_{3}=\int_{0}^{t}dt_{1}\langle\hat{\Gamma}(t)\hat{\Gamma}^{\dagger}(t_{1})\rangle e^{i\Omega(t-t_{1})}, \\
~I_{4}=\int_{0}^{t}dt_{1}\langle\hat{\Gamma}^{\dagger}(t)\hat{\Gamma}(t_{1})\rangle e^{i\Omega(-t+t_{1})},
\end{cases}
\end{equation}

\begin{equation}
\langle\hat{A}\hat{B}\rangle\equiv\mathrm{Tr}_{\mathcal{H}_{b}} \; (\hat{A}\hat{B}\hat{\rho}_{b}).
\end{equation}

We suppose the form of the correlation function to be 
\begin{equation}
\langle\hat{b}_{\omega_{1}}\hat{b}_{\omega_{2}}\rangle\equiv2\pi M(\omega_{1})\delta(2\Omega-\omega_{1}-\omega_{2}), ~\mathrm{and}~
\langle\hat{b}^{\dagger}_{\omega_{1}}\hat{b}_{\omega_{2}}\rangle\equiv2\pi N(\omega_{1})\delta(\omega_{1}-\omega_{2}),
\end{equation}
where $N(\omega_{1})=\hat{b}_{\omega_1}^{\dagger}\hat{b}_{\omega_1}$, then $I_{1}$ become
\begin{equation}
I_{1}(t)=2\pi\left(i\hbar\right)^{2}\int_{0}^{t}d\tau\int_{-\infty}^{\infty} d(\Delta\omega) \; \kappa(\Omega+\Delta\omega)  \kappa(\Omega-\Delta\omega)M(\Omega+\Delta\omega)e^{-i\Delta\omega\tau},
\end{equation}
where $\Delta\omega=\omega-\Omega$.

Here we assume that $\kappa(\omega)$, $M(\omega)$ are a function
that changes slowly around $\omega=\Omega$. Therefore we can take the Markov approximation and $I_{1}$ becomes
\begin{equation}
I_{1}\approx2\pi\left(i\hbar\right)^{2}\int_{0}^{t}d\tau\int_{-\infty}^{\infty} d(\Delta\omega) \; \kappa(\Omega+\Delta\omega)^{2}M(\Omega+\Delta\omega)e^{-i\Delta\omega\tau}.
\end{equation}
Here we assume that the integrals for frequency decay rapidly with
respect to time, take the time integrals to infinity, and interchange
the order of the integrals, then we obtain
\begin{equation}
\begin{aligned}I_{1}\approx & 2\pi\left(i\hbar\right)^{2}\int_{-\infty}^{\infty}d(\Delta\omega) \; \kappa^{2}(\Omega+\Delta\omega)M(\Omega+\Delta\omega)(-i) \left[\mathcal{P}\frac{1}{\Delta\omega}+i \pi\delta(\Delta\omega) \right],\\
\equiv & 2\left(i\hbar\right)^{2}\gamma M(\Omega)+i\bar{\Delta}
\end{aligned}
\end{equation}
where $\mathcal{P}$ is the principal value of Cauchy and $\gamma\equiv\kappa^{2}(\Omega)$.

In the following, we ignore $\bar{\Delta}$ as having a small effect.

Similarly, we obtain
\begin{eqnarray}
\begin{cases}
~I_{1}\approx2\left(i\hbar\right)^{2}\gamma M(\Omega), \\
~I_{2}\approx2(i\hbar)^{2}\gamma M^{*}(\Omega), \\
~I_{3}\approx2(i\hbar)^{2}\gamma\left(N(\Omega)+1\right), \\
~I_{4}\approx2(i\hbar)^{2}\gamma N(\Omega), \end{cases}
\end{eqnarray}
where $N(\Omega)=\hat{a}^{\dagger}\hat{a}$ is the number operator of photon in the cavity.

After all, including the part of $\hat{H}_{Cavity,I}$, the master
equation becomes
\begin{equation}
\begin{aligned}\frac{d}{dt}\hat{\rho}(t) & \approx  [\hat{H}_{Cavity,I},\hat{\rho}]\\
 & +\frac{\gamma}{2}\left(N(\Omega)+1\right)\left(2\hat{a}\hat{\rho}\hat{a}^{\dagger}-\hat{a}^{\dagger}\hat{a}\hat{\rho}-\hat{\rho}\hat{a}^{\dagger}\hat{a}\right)+\frac{\gamma}{2}N(\Omega)\left(2\hat{a}^{\dagger}\hat{\rho}\hat{a}-\hat{a}\hat{a}^{\dagger}\hat{\rho}-\hat{\rho}\hat{a}\hat{a}^{\dagger}\right)\\
 & +\frac{\gamma}{2}M(\Omega)\left(2\hat{a}^{\dagger}\hat{\rho}\hat{a}^{\dagger}-\hat{a}^{\dagger}\hat{a}^{\dagger}\hat{\rho}-\hat{\rho}\hat{a}^{\dagger}\hat{a}^{\dagger}\right)+\frac{\gamma}{2}M^{*}(\Omega)\left(2\hat{a}\hat{\rho}\hat{a}-\hat{a}\hat{a}\hat{\rho}-\hat{\rho}\hat{a}\hat{a}\right).
\end{aligned}
\label{eq:master_MN-1}
\end{equation}
When the heat bath is characterized by a temperature $T$, we choose
\begin{equation}
M(\Omega)=0, ~
N(\Omega)=\frac{1}{e^{\beta\hbar\Omega}-1}.
\end{equation}

In the following, we will consider the case of $T=0$. Then, the master equation becomes simple and tractable,
\begin{equation}
\begin{aligned}\frac{d}{dt}\hat{\rho}(t)= & [\hat{H}_{Cavity,I},\hat{\rho}]+\frac{\gamma}{2}\left(\hat{a}\hat{\rho}\hat{a}^{\dagger}-\hat{a}^{\dagger}\hat{a}\hat{\rho}-\hat{\rho}\hat{a}^{\dagger}\hat{a}\right)\end{aligned}.
\end{equation}

To estimate this master equation, it is known to be useful the Glauber-Sudarshan's P representation \cite{P-rep} in terms of coherent state:  \begin{equation}
\hat{\rho}(t)=\int P(\alpha)|\alpha\rangle\langle\alpha|d^{2}\alpha,
\end{equation}

Then, the master equation becomes
\begin{equation}
\frac{\partial}{\partial t}P(\alpha)=H_{Cavity}P(\alpha)+\frac{\gamma}{2}\left(\frac{\partial}{\partial\alpha^{*}}\alpha^{*}+\frac{\partial}{\partial\alpha}\alpha\right)P(\alpha)
\end{equation}

If we take the second-order squeezing term as a term in $H_{Cavity}$,
we obtain the Fokker-Planck equation
\begin{equation}
\begin{aligned}\frac{\partial}{\partial t}P(\alpha)= & \left\{ -\left(\lambda^{*}(2\Omega)\frac{\partial}{\partial\alpha^{*}}\alpha+\lambda(2\Omega)\frac{\partial}{\partial\alpha}\alpha^{*}\right)\right.\\
 & \left.+\frac{\gamma}{2}\left(\frac{\partial}{\partial\alpha^{*}}\alpha^{*}+\frac{\partial}{\partial\alpha}\alpha\right)+\frac{1}{2}\left(\lambda^{*}(2\Omega)\frac{\partial^{2}}{\partial\alpha^{*2}}+\lambda(2\Omega)\frac{\partial^{2}}{\partial\alpha^{2}}\right)\right\} P(\alpha).
\end{aligned}
\label{eq:fockerplanck}
\end{equation}

Generally, Fokker-Planck equation reads 
\begin{equation}
\frac{\partial}{\partial t}P(t)=\left\{ \frac{\partial}{\partial\alpha_{i}}A_{ij}\alpha_{j}+\frac{1}{2}\frac{\partial}{\partial\alpha_{i}}\frac{\partial}{\partial\alpha_{i}}B_{ij}B_{ji}+ (c.c.) \right\} P(\alpha)
\end{equation}

can be transformed into the Langevin equation
\begin{equation}
\frac{d\alpha_{i}}{dt}=-A_{ij}\alpha_{j}+B_{ij}\xi^{j},
\end{equation}
and its complex conjugate equation, where $\xi^{j}$ is a Gaussian noise.

When $A_{ij}$ is diagonalizable and its eigenvalue is real and positive,
then we obtain steady-state solution
\begin{equation}
\alpha_{SS}^{i}(t)=\int_{-\infty}^{t}\left(e^{-(t-t')A}\right)_{ij}B_{jk}\xi^{k}
\end{equation}

So correlation function
\begin{equation}
G_{SS}(t-s)\equiv\langle\alpha_{SS}^{i}(t),\alpha_{SS}^{Tj}(s)\rangle=\int_{-\infty}^{min(t,s)}e^{-(t-t')A}BB^{T}e^{-(s-t')A^{T}}dt'
\end{equation}

is the Fourier transform of the spectrum matrix
\begin{equation}
S(\omega)=\frac{1}{2\pi}\int_{-\infty}^{\infty}e^{-i\omega\tau}G_{SS}(\tau)d\tau=\frac{1}{2\pi}\left(A+i\omega\right)^{-1}BB^{T}\left(A^{T}-i\omega\right)^{-1}.
\end{equation}

In case of Eq.(\ref{eq:fockerplanck}),
\begin{equation}
A=\left(\begin{array}{cc}
\frac{\gamma}{2} & -\lambda(2\Omega)\\
-\lambda^{*}(2\Omega) & \frac{\gamma}{2}
\end{array}\right), ~
B=\left(\begin{array}{cc}
\lambda(2\Omega) & 0\\
0 & \lambda^{*}(2\Omega)
\end{array}\right)
\end{equation}
we can predict the spectrum $S_{i}^{OUT}$, where $i=1, 2$ correspond to $\theta=0, \pi/2$, respectively, as follows:
\begin{eqnarray}
&&S_{1}^{OUT}=1+\frac{2\gamma|\lambda(2\Omega)|^{2}}{\left(\frac{\gamma}{2}-|\lambda(2\Omega)|\right)^{2}+\left(\omega-\Omega\right)^{2}},\label{eq:SOUT1} \\
&&S_{2}^{OUT}=1-\frac{2\gamma|\lambda(2\Omega)|^{2}}{\left(\frac{\gamma}{2}+|\lambda(2\Omega)|\right)^{2}+\left(\omega-\Omega\right)^{2}}.\label{eq:SOUT2} 
\end{eqnarray}
The same result was derived in the input-output theory \cite{key-2, key-4}.  Therefore, the input-out-put theory is within the category of Fokker-Planck equation.

If $\hat{H}_{Cavity}$ contains a higher nonlinear term, the higher-order
derivatives with respect to $\alpha$ and $\alpha^*$ arise, so that in this case the equation is beyond the Fokker-Planck equation which is the second order differential equation. 

From this we may understand that the Fokker-Planck equation is a powerful method for the dissipative system with quadratic Hamiltonian, but we will confront with a difficulty for the Hamiltonian with higher order terms than quadratic.

The essential place which we have to apply the Fokker-Planck equation is the laser oscillation \cite{Fokker-Planck}, where the infinite number of photons are involved; in the process of the laser pumping or the excitation of atoms, and the subsequent stimulated and spontaneous emissions of photons, under the various sources of absorption.  

\subsection{Relation of our paper to this Appendix B}
Our paper uses the solution of Fokker-Planck equation applied for the fluctuation of phase $\theta$ of photons, defined by $\alpha=|\alpha| e^{i \theta}$.  

The Fokker-Planck equation of this problem is \cite{Fokker-Planck},
\begin{eqnarray}
\frac{\partial}{\partial t} P(\alpha, t)= -\frac{1}{2} \left\{\frac{\partial}{\partial \alpha} \left[ (A-C-B |\alpha|^2 ) \alpha P(\alpha, t)\right] + (c.c.) \right\} + A \frac{\partial^2}{\partial \alpha \partial \alpha^*} P(\alpha, t).
\end{eqnarray}

Let $b$, $a$ denote the upper level and the lower level of atoms, between which the emission and absorption of light occurs.  Atoms at the ground state $c$ are pumped to the higher level $b$, but the atom at $b$ and $a$ can loose energy and be relaxed to the ground $c$ by rates $\gamma_b$ and $\gamma_a$, respectively.  The various constants are given by
\begin{eqnarray}
&&A=\frac{N_a rg}{2 \gamma_b}, \; B=\frac{C}{\beta}, \; C=\frac{c}{2L} T+ \alpha(\omega) c, 
\end{eqnarray}
where $N_a$ is the number of atoms, $r$ is the rate of pumping per time, $c$ is light velocity, $L$ and $V$ are length and volume of the Fabry-P\'erot cavity, $T$ is transmission coefficient of the mirror, $\alpha(\omega)$ is absorption rate per length by the matter inside cavity.  The $g$ and $\beta$ are given by the dipole emission rate $| e\bm{r}_{ab}|^2$ of photon from the atom, 
\begin{eqnarray}
g=2 \pi \frac{e^2 \omega_{ba} | \bm{r}_{ab}|^2}{3 \epsilon_0 V \hbar (\gamma_a+\gamma_b)},\;  \beta=\frac{2 \gamma_a\gamma_b}{(\gamma_a+\gamma_b)g}.
\end{eqnarray}

In terms of polar coordinate, $\alpha=|\alpha|e^{i\theta}$, we have
\begin{eqnarray}
&&\frac{\partial}{\partial t} P(|\alpha|, \theta, t)=- \frac{1}{2|\alpha|}\frac{\partial}{\partial |\alpha|} \left[ |\alpha|^2 (A-C-B |\alpha|^2 ) P(|\alpha|, \theta, t)\right] \nonumber \\
&&~~~~~~~~~~~~~~~~~~+\left(\frac{A}{4|\alpha|^2} \frac{\partial^2}{\partial \theta^2} + \frac{A}{4|\alpha|} \frac{\partial}{\partial |\alpha|} |\alpha|\frac{\partial}{\partial |\alpha|}\right) P(|\alpha|, \theta, t).
\end{eqnarray}
Assume the solution is $P(|\alpha|) \times P(\theta, t)$, then we have two equations,
\begin{eqnarray}
\begin{cases}
~- \frac{1}{2|\alpha|}\frac{\partial}{\partial |\alpha|} \left[ |\alpha|^2 (A-C-B |\alpha|^2 ) P(|\alpha|)\right] +\left(\frac{A}{4|\alpha|} \frac{\partial}{\partial |\alpha|} |\alpha|\frac{\partial}{\partial |\alpha|}\right) P(|\alpha|) =\lambda P(|\alpha|), \\
~\frac{\partial}{\partial t} P(\theta, t)=\left(\frac{A}{4|\alpha|^2} \frac{\partial^2}{\partial \theta^2}-\lambda \right) P(\theta, t).
\end{cases}
\end{eqnarray}
The first equation gives the stationary distribution for the photon number, $n=|\alpha|^2$, while the second equation describes the temporal development of the phase $\theta$.  For simplicity we take $\lambda=0$.  Then the solution of the first equation is 
\begin{eqnarray}
P(|\alpha|) \propto e^{ -\frac{B}{2A} |\alpha|^4 + \frac{A-C}{A} |\alpha|^2}
\propto e^{ -\frac{B}{2A} \left(|\alpha|^2 - \frac{A-C}{B} \right)^2}
\end{eqnarray} 

This Gaussian distribution gives the averaged number of photon and its standard deviation as
\begin{eqnarray}
\langle n \rangle = \frac{A-C}{B}, ~~ \sigma_n= \sqrt{\frac{A}{B}}.
\end{eqnarray}

The laser is above the threshold, when the pumping rate $A$ is much larger than the cavity loss $C$, $A \gg C$, leading to $\langle n \rangle=\sigma_n^2$, that is the laser light above threshold obeys the Poisson distribution, and the two point correlation function becomes unity:
\begin{eqnarray}
g_{12}=\frac{\langle \bm{E}_1^2\bm{E}_2^2 \rangle}{\langle \bm{E}_1^2 \rangle \langle \bm{E}_2^2 \rangle}=1.
\end{eqnarray}
This means that the laser light becomes classical light.

As for the second equation, if choosing $|\alpha|^2$ as the average number of photons $\langle n \rangle$, it becomes a diffusion equation with a diffusion constant $D=A/2\langle n \rangle$, (Einstein's equation for Brownian motion):  
Then, we have
\begin{eqnarray}
\frac{\partial}{\partial t} P(\theta, t)=\frac{A}{4\langle n \rangle} \frac{\partial^2}{\partial \theta^2} P(\theta, t).
\end{eqnarray}
The solution is 
\begin{eqnarray}
P(\theta, t) = \sqrt{\frac{\langle n \rangle}{\pi A t}} \times  e^{-\frac{ \langle n \rangle (\theta-\theta_0)^2}{At}},
\end{eqnarray} 
for which the standard deviation squared increases linearly in time, but can be zero: 
\begin{eqnarray}
\sigma_{\theta}^2(t)= \frac{A}{2\langle n \rangle}t = D t \approx \frac{B}{2} t \ll 1,  
\end{eqnarray}
for $A \gg C$ above threshold.  If an actual laser gives $(A, C)=O(10^6)/$[s], and $B \sim 0.1/$[s], $\sigma_{\theta}^2(t) \sim 10^{-9} \ll 1$, when the diffusion time 10 [s] is compared with the typical time of laser operation, $t=2L/c= 10^{-8}$[s] for the cavity with length $L=1.5$ [m].

Now, we can understand that if the laser is tuned sufficiently above the threshold $A \gg C$, the laser light becomes the classical one, having a Poisson distribution for the number, and a common phase which is constant without caused by the diffusion. 
This is the starting line of our paper, by assuming the classical electric field for the input laser light.

\section*{Appendix C: QED in non-uniform dielectric matter}
QED in non-uniform dielectric matter is studied by L. Kn\"oll, S. Scheel, D-G. Welsch \cite{QED}. For the recent developments, refer to the book by P. D. Drummond and M. Hillery \cite{QED}. 

Considering a system of doped atoms and radiation field (photons) in the dielectric matter.  The dielectric matter is assumed to consist of a finite number of dipoles $\bm{X}_i$ with discrete spectra $\omega_i$ $(i=1-N$), and the infinite number of dipoles $\bm{X}_{\omega}$ with continuum spectra $\omega$ $(0 \le \omega < \infty)$.  This is the microscopic Hopfield model of dielectrics. The Lagrangian is 
\begin{eqnarray}
\mathcal{L}=\mathcal{L}_{atom}+\mathcal{L}_{rad}+\mathcal{L}_{matters}+\mathcal{L}_{int},
\end{eqnarray}
where
\begin{eqnarray}
\begin{cases}
~\mathcal{L}_{rad}=\frac{1}{2} \varepsilon_0 \left(\bm{E}^2 - c^2 \bm{B}^2 \right), \; 
~\mathcal{L}_{atom}=\frac{m_a}{2}\dot{\bm{x}}_a^2, \\
~\mathcal{L}_{matters}= \sum_{i=1}^N \frac{\mu}{2} \left( \dot{\bm{X}}_i^2 -\omega_i^2 \bm{X}_i^2 \right)
+\int d\omega \; \frac{\mu}{2} \left( \dot{\bm{X}}_{\omega}^2 -\omega^2 \bm{X}_{\omega}^2 \right) - \int d\omega \; \dot{\bm{X}}_{\omega} v_i(\omega) \bm{X}_i, \\
~\mathcal{L}_{int}=-e_a \left(\dot{\bm{x}}_a \bm{A}(x_a)-\phi(x_a)\right) -  \sum_{i=1}^N \alpha_i \left(\dot{\bm{X}}_i \bm{A}(\bm{X}_i) - \phi(\bm{X}_i) \bm{\nabla} \bm{X}_i \right).
\end{cases}
\end{eqnarray}

Putting aside the Lagrangian of the doped atom, the electric field $\bm{E}$ couples to the discrete dipoles $\bm{X}_i$, but not to the continuum dipoles $\bm{X}_{\omega}$, from which we understand in the first order perturbation theory, the effective action for photons becomes,
\begin{eqnarray}
&&S_{eff}=\int d^3r \; \frac{1}{2} \varepsilon_0 \left(\bm{E}^2 - c^2 \bm{B}^2 \right) +\int d^3r \int d^3r' \; \frac{1}{2} \bm{E}(r) \sum_{i=1}^N \alpha_i^2 \langle \bm{X}_i(r) \bm{X}_i(r') \rangle \bm{E}(r') \nonumber \\
&&+ \int d^3r \int d^3 r' \int d\omega \; \bm{A}_{\omega}(r) \sum_{i=1}^N \alpha_i \langle \dot{\bm{X}}_i(r) \dot{\bm{X}}_i(r') \rangle v_i(\omega)  \bm{X}_{\omega}(r')\\
&&=\int d^3r \; \frac{1}{2} \varepsilon (r) \left(\bm{E}(r)^2 - c'(r)^2 \bm{B}^2 \right) + \int d^3r \; \bm{A}_{\omega}(r)\bm{j}_{\omega}(r), 
\end{eqnarray}
where $\langle \bm{X}_i(r) \bm{X}_i(r') \rangle$ is a two point Green function or the Feynman propagator, and
\begin{eqnarray}
\varepsilon=\epsilon_0(1+\chi^{(0)}), ~\mathrm{and}~\bm{j}_{\omega}(r)=\sum_{i=1}^N \alpha_i \langle \dot{\bm{X}}_i(r) \dot{\bm{X}}_i(r') \rangle v_i(\omega)  \bm{X}_{\omega}(r').
\end{eqnarray}

In the above discussion, the current $\bm{j}_{\omega}(r)$ is induced by the continuum modes of the dipoles, $\bm{X}_{\omega}(r)$, which is considered in \cite{QED}, the noise of electric current coming from the reservoir, being understood by the Fokker-Planck equation. 

The effective action gives the diversion and the absorption of photon, since $\varepsilon$ is a complex number having real part and imaginary part, so that the refraction constant $n=\sqrt{\varepsilon}$ is also complex number.  In this second order effective action of photon, the propagator of photon in two dimensional space-time reads
\begin{eqnarray}
G_{\omega}(r, t; r', t) \propto e^{-i \left(\omega(t-t')- \frac{n_R}{c} \omega |z-z'|\right)}e^{-\frac{n_I}{c} \omega |z-z'|}. \label{propagator in the medium}
\end{eqnarray}

As a microscopic model of dielectrics, this simple model of dipoles \cite{QED} is further developed, using multi-polar expansion and the polariton, considering the levels of $N$ dielectric atoms as the levels of a higher spin $\frac{N}{2}$, and so on.  

For the non-linear QED, a delicate problem seems to be discussed in the process of quantization \cite{QED}.  To consider this problem, we have to think about carefully, the difference of the original action and the non-linear effective action derived from it, without doubling the quantization procedures.  For example, we are assumed to start from an action $S(0)$, perform the quantization (or path integration) over the higher energy modes with $E > \mu$, and to obtain an non-linear effective action $S(\mu)$ at energy scale $\mu$.  Then, the further quantization (or path integration) is allowed only for the lower energy modes of $E < \mu$.

\subsection{Relation of our paper to this Appendix C}
Our paper does not consider the microscopic structure of non-linear material, but the expression of dielectric permeability of it phenomenologically, using the values $\chi^{(n)}$ obtained by measurement.  The purpose of out paper is to examine how the perturbation theory works well even in the presence of non-uniform dielectric matter and mirrors.  Therefore, the quantization is performed for the quasi-free Hamiltonian (the quadratic part in the non-linear QED), and the real interaction Hamiltonian is considered as a perturbation.  This is a kind of ``interaction picture'', in which the creation and annihilation operator is defined for the quasi-free Hamiltonian, and the interaction Hamiltonian is expressed in terms of the operators.  Therefore, the numerator of our propagators in Sec.4.2 coincides with Eq.(\ref{propagator in the medium}) in \cite{QED}.
As far as the perturbation theory is applied as ours, no delicate problem discussed in \cite{QED} appears.

 \section*{Appendix D: -Generalized squeezing-}
On the normalizablity of the generalized squeezing state, Fisher, Nieto and Sandberg \cite{key-41} shows that the vacuum expectation value of the generalized squeezed state diverges and the perturbative calculations are broken.
Elyutin and Klyshko \cite{key-42} indicates that the state diverges
in a finite time and behaves very differently from the normal squeezing
state. To get around this problem, k-photon generalized boson operator was studied \cite{key-50,key-51} which uses the following operators \cite{key-52}:
\begin{eqnarray}
\hat{A}_{(k)}=\left(\left[\left[\frac{\hat{n}}{k}\right]\right]\frac{\left(\hat{n}-k\right)!}{\hat{n}!}\right)^{1/2}\left(\hat{a}^{\dagger}\right)^{k}, \; \mathrm{and}\; \hat{S}_{(k)}=\exp\left(z\hat{A}_{(k)}^{\dagger}-z^{*}\hat{A}_{(k)}\right),
\end{eqnarray}
where
\begin{eqnarray}
\hat{n}=\hat{a}^{\dagger}\hat{a},\; [\hat{A}_{(k)},\hat{A}_{(k)}^{\dagger}]=1,\; [\hat{n},\hat{A}_{(k)}]=-k\hat{A}_{(k)},
\end{eqnarray}
and $\left[\left[\frac{\hat{n}}{k}\right]\right]$ denotes the maximal integer less than or equal to $\frac{\hat{n}}{k}$.

On the other hand, Braunstein and McLachlan \cite{generalized squeezing} argues that the divergence of the vacuum expectation value is a mere mathematical artifact, and compute it numerically using the Pade approximation. Later, Braunstein and Caves \cite{key-45} calculates
statistics on $\hat{A}_k$ by homodyne and heterodyne measurements. Banaszek
and Knight \cite{key-46} computes the Wigner function about $k=3$
and shows that there is a negative region in its value. Govia, Pritchett
and Wilhelm \cite{key-47} proposes a method to generate generalized
squeezed states from coherent states using Josephson photomultiplier
(JPM).

Experimentally, Cooper et al. \cite{key-48} generated a multiphoton
Fock state. Recently, Chang {\it et al.} \cite{key-49} observed the distribution of star states, which have been expected to be observed for $k=3$. 

\end{document}